\documentclass[reprintamsfonts,amssymb,amsmath,floatfix]{revtex4}

\usepackage[dvips]{graphicx}
\usepackage{mathrsfs}
\usepackage{amsmath,amssymb}
\usepackage{hhline}
\usepackage{dcolumn}
\usepackage{bm}

\usepackage[dvips]{color}

\newcommand{\tf}{\tilde{f}}
\newcommand{\tu}{\tilde{u}_{f}}
\newcommand{\tv}{\tilde{v}_{f}}
\newcommand{\mgq}{\mathcal{G}_q}
\newcommand{\mgr}{\mathcal{G}_r}

\newcommand{\be}{\begin{equation}}
\newcommand{\ee}{\end{equation}}
\newcommand{\bi}{\begin{itemize}}
\newcommand{\ei}{\end{itemize}}

\begin{document}

\title{Structure of percolating clusters in random clustered networks}
\date{\today}

\author{Takehisa Hasegawa}\email{takehisa.hasegawa.sci@vc.ibaraki.ac.jp}
\author{Shogo Mizutaka\footnote{Present address: School of Knowledge Science, Japan Advanced Institute of Science and Technology, 1-1 Asahidai, Nomi 924-1292, Japan}}\email{mizutaka@jaist.ac.jp}
\affiliation{Department of Mathematics and Informatics, Ibaraki University, 2-1-1 Bunkyo, Mito, Japan 310-8512}

\begin{abstract}
We examine the structure of the percolating cluster (PC) formed by site percolation on a random clustered network (RCN) model.
Using the generating functions, we formulate the clustering coefficient and assortative coefficient of the PC. 
We analytically and numerically show that the PC in the highly clustered networks is clustered even at the percolation threshold.
The assortativity of the PC depends on the details of the RCN. 
The PC at the percolation threshold is disassortative when the numbers of edges and triangles of each node are assigned by Poisson distributions, but assortative when each node in an RCN has the same small number of edges, most of which form triangles.
This result seemingly contradicts the disassortativity of fractal networks, although the renormalization scheme unveils the disassortative nature of a fractal PC.
\end{abstract}

\maketitle

\section{Introduction}
A number of studies on complex networks have reported the structural characteristics of a real network ranging from the World Wide Web to food webs \cite{barabasi2016network,newman2018networks,pastor2015epidemic,dorogovtsev2008critical}. 
Numerous real networks are scale-free, i.e., the distribution $p_k$ of degree $k$ obeys a power law. 
Most of the real networks are small world, indicating that the mean shortest path length scales with the logarithm of the number of nodes and the clustering coefficient, which is the mean probability that two randomly-chosen neighbors of a randomly-chosen node are adjacent, is high.
Real networks would be classified by the degree-degree correlation, i.e., the correlation between the degrees of directly connected nodes. 
Social networks have a positive degree-degree correlation in which similar degree nodes are more likely to connect to each other while biological and technological networks have a negative degree-degree correlation indicating that dissimilar degree nodes are more likely to connect to each other.
Furthermore, Song et al. reported on the fractality of real networks \cite{song2005self,song2006origins,song2007calculate}: 
some real networks, such as the World Wide Web and protein-protein interaction networks, are fractal in the sense that the number of boxes for tiling a network decreases with the radius of boxes in a power law manner.

It is crucial to understand how structural characteristics are related to each other.
Yook et al. \cite{yook2005self} discovered from real network data that fractal networks have a negative degree-degree correlation, namely, disassortativity.
This empirical rule is observed in the synthetic models of fractal networks \cite{song2006origins,rozenfeld2007fractal}, critical branching trees \cite{kim2009disassortativity}, and connected components at a critical state of an uncorrelated network model \cite{tishby2018revealing,bialas2008correlations}. 
Furthermore, there are related works concerning the degree-degree correlation of spanning trees in fractal and small-world networks \cite{goh2006skeleton,kim2004scale,wei2016emergence} and the converse condition that disassortativity makes a network fractal \cite{fujiki2017fractality}. 
However, it still remains unclear why fractal networks possess disassortativity and how robust the empirical rule is. 

Site percolation on networks is known to exhibit a phase transition concerning clusters, which are connected components of occupied nodes. 
When the number of nodes is sufficiently large, the largest cluster is small and finite for $f<f_c$; it occupies a finite fraction of the whole network and is called the {\it percolating cluster} (PC) for $f>f_c$; and it is a fractal at $f=f_c$ \cite{stauffer1994introduction}. 
Here $f$ is a fraction of the occupied nodes and $f_c$ is called the percolation threshold.
The analysis of the largest cluster at $f=f_c$, which is called the fractal PC, leads us to further examine the relation between the fractality and the disassortativity in complex networks.

In a percolation process, a network splits into multiple connected components.
It should be noted that the structural properties of a connected component are different from those of the whole network if the network is not singly connected \cite{bialas2008correlations,tishby2018revealing,tishby2018generating,tishby2018statistical,mizutaka2018disassortativity}.
Recent studies have focused on the methods to extract the infinitely large connected component from uncorrelated networks and compute its properties (e.g., degree distribution $p_k$, average degree $\bar{k}_{\rm nn}(k)$ of nodes adjacent to degree $k$ nodes \cite{bialas2008correlations}, and assortative coefficient $r$ defined by Pearson's correlation coefficient for degrees of directly connected nodes \cite{tishby2018revealing}).
Previous work \cite{mizutaka2018disassortativity} considered a PC formed by site percolation on uncorrelated networks and investigated the properties of the PC.
For uncorrelated random networks obeying an arbitrary degree distribution with a finite third moment, the PC possesses a disassortativity above the percolation threshold \cite{mizutaka2018disassortativity}: the assortative coefficient $r$ is always less than zero. 
Moreover, the average degree $\bar{k}_{\rm nn}(k)$ of the nodes adjacent to the degree $k$ nodes is proportional to $k^{-1}$ at $f \to f_{\rm c}$. 
These indicate that the fractal PC is disassortative when it is formed by site percolation on uncorrelated networks.

The present study is a continuation of our previous work \cite{mizutaka2018disassortativity} and discusses whether the disassortativity of PCs is established in correlated networks.
It is also interesting how other structural properties of the PC differ from those of original networks: e.g., how is the PC in a clustered network clustered?
Newman~\cite{newman2009random} and Miller~\cite{miller2009percolation} independently introduced a random graph model with clustering, many of whose network properties can be well described via a generating function analysis.
This model is highly clustered and even assortative (as shown below)---it is suitable for discussing the above-mentioned question.
In this study, we consider site percolation on the random clustered network (RCN) model to investigate the structural properties, the clustering coefficient and the assortative coefficient, of the PC.
Our generating function analysis describes the structure of the PC well: it perfectly agrees with the corresponding Monte Carlo simulations.
We show that the PC formed by site percolation in highly clustered networks is clustered even at the percolation threshold.
With respect to the assortativity, both analytical estimates and simulation results seemingly contradict the disassortativity of fractal networks: the fractal PC is assortative when the nodes in an RCN have the same small number of edges, most of which form triangles.
Our discussion focuses on why a positive assortativity is observed in a fractal PC.

The paper is organized as follows. 
In Sec.~\ref{sec:model}, we introduce the RCN model and recall the generating function approach for deriving its clustering coefficient and assortative coefficient.
In Sec.~\ref{sec:analysis}, we analyze the structure of the PC formed by site percolation on RCNs. 
We derive the clustering coefficient (Sec.~\ref{sec:cluster}) and assortative coefficient (Sec.~\ref{sec:assortativity}) of the PC, 
applying our analysis to two types of RCNs (Sec.~\ref{sec:numericalcheck}) in order to investigate the structures of the PC.
In Sec.~\ref{sec:discussion}, we discuss the robustness of the fractal PC's disassortativity.

\section{Random clustered network model \label{sec:model}}

\begin{figure}[b]
\begin{center}
\includegraphics[width=.25\textwidth]{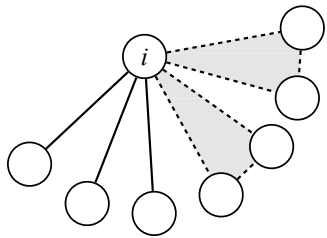}
\caption{
Node $i$ in this figure has 3 single edges and 2 triangles. The solid lines and dashed lines represent the single edges and triangle edges, respectively. Triangles are filled in with gray. The degree of node $i$ is $k_i=3+2\times2=7$. 
}
\label{fig:edges}
\end{center}
\end{figure}

The RCN model introduced by Newman~\cite{newman2009random} generalizes the configuration model to incorporate clustering. 
We assume that the joint probability, $p_{s,t}$, for the mean fraction of nodes with $s$ single edges and $t$ triangles is given and assign $s_i$ {\it edge stubs} and $t_i$ {\it triangle stubs} to each node $i$ according to $p_{s,t}$ under the constraint that $\sum_{i} s_i$ and $\sum_{i} t_i$ are multiples of 2 and 3, respectively. 
Given these stubs, we create a network by randomly selecting pairs of edge stubs and joining them to make {\it single edges} and by randomly selecting triples of triangle stubs and joining them to form triangles whose edges are referred to as {\it triangle edges}. 
This results in a random network in which the number of single edges incident to each node and the number of triangles it participates in are distributed according to $p_{s,t}$.
Note that the total degree $k$ of a node with $s$ single edges and $t$ triangles is $k=s+2t$  (Fig.~\ref{fig:edges}).

The clustering coefficient $C_0$ of the RCN is given by the generating functions~\cite{newman2009random}. 
First, we introduce the generating function $G_p(x,y)$ for the joint probability $p_{s,t}$,
\be
G_p(x,y)=\sum_{s=0}^\infty \sum_{t=0}^\infty p_{s,t} x^s y^t.
\ee
Because the full degree distribution $p_k$ is written as $p_k=\sum_{s,t} p_{s,t} \delta_{k, s+2t}$ using the Kronecker delta $\delta_{ij}$, the generating function $G_{\rm tot}(z)$ for the full degree distribution $p_k$ is presented as follows:
\be
G_{\rm tot}(z) = \sum_{k=0}^\infty p_k z^k = G_p(z,z^2).
\ee
The average degree $\langle k \rangle$ is obtained from $G_{\rm tot}(z)$ as follows:
\be
\langle k \rangle = \frac{\partial G_{\rm tot}(z)}{\partial z} \Big|_{z=1} = \langle s \rangle + 2 \langle t \rangle, 
\ee
where $\langle s \rangle = \sum_{s,t} s p_{s,t}$ and $\langle t \rangle = \sum_{s,t} t p_{s,t}$.
For the RCN with $N$ nodes, the number $N_3$ of the connected triplets and the number $N_\Delta$ of the triangles are given by the generating functions $G_p(x,y)$ and $G_{\rm tot}(z)$ \cite{newman2009random}:
\be
N_3 = N \sum_k \frac{k(k-1)}{2} p_k = \frac{1}{2} N \frac{\partial^2 G_{\rm tot}(z)}{\partial z^2} \Big|_{z=1},
\ee
and 
\be
3 N_\Delta = N \sum_{s,t} t p_{s,t} = N \frac{\partial G_p(x,y)}{\partial y} \Big|_{x=y=1}.
\ee
We can then write the clustering coefficient $C_0$ of the RCN as follows:
\be
C_0 = \frac{3 N_\Delta}{N_3} = 2 \frac{\partial G_p(x,y)}{\partial y} \Big|_{x=y=1} \Big/ \frac{\partial^2 G_{\rm tot}(z)}{\partial z^2} \Big|_{z=1}. \label{eq:clusteringCoefficient}
\ee

Moreover, the assortative coefficient $r_0$ of the RCN is formalized using the generating functions. 
We consider two types of excess degree distributions \cite{newman2009random}: 
$q_{s,t}$, which is the probability that a node reached by traversing a single edge has $s+1$ single edges and $t$ triangles, and $r_{s,t}$, which is the probability that a node reached by traversing a triangle has $s$ single edges and $t+1$ triangles.
These probabilities are naturally derived as follows:
\be
q_{s,t}=\frac{s+1}{\langle s \rangle} p_{s+1,t}
\ee
and
\be
r_{s,t}=\frac{t+1}{\langle t \rangle} p_{s,t+1}.
\ee
We then introduce the generating functions for $q_{s,t}$ and $r_{s,t}$ as
\be
G_q(x,y)=\sum_{s=0}^\infty \sum_{t=0}^\infty q_{s,t} x^s y^t=\frac{1}{\langle s \rangle} \frac{\partial G_p(x,y)}{\partial x}
\ee
and
\be
G_r(x,y)=\sum_{s=0}^\infty \sum_{t=0}^\infty r_{s,t} x^s y^t=\frac{1}{\langle t \rangle} \frac{\partial G_p(x,y)}{\partial y},
\ee
respectively. 
Here we denote the probability of choosing a single edge by
\be
P_s = \frac{\langle s \rangle}{\langle s \rangle+2\langle t \rangle},
\ee
and the probability of choosing a triangle edge by
\be
P_t = 1-P_s = \frac{2 \langle t \rangle}{\langle s \rangle+2\langle t \rangle}.
\ee
Introducing the probability $Q_0(k, k')$ that two ends of a randomly chosen edge have degrees $k+1$ and $k'+1$ and the probability $Q_0(k) [=\sum_{k'}Q_0(k,k')]$ that an edge reaches a node with degree $k+1$, we can calculate the assortative coefficient $r_0$ of the RCN from the following equation:
\be
r_0=\frac{\partial_x \partial_y B_0(x,y) - (\partial_x S_0(x))^2}{(x \partial_x)^2 S_0(x) - (\partial_x S_0(x))^2} \Big|_{x=y=1}, \label{eq:r0}
\ee
where $B_0(x,y)$ is the generating function for $Q_0(k_1,k_2)$,
\be
B_0(x,y) = \sum_{k_1=0}^\infty \sum_{k_2=0}^\infty Q_0(k_1,k_2) x^{k_1} y^{k_2} = P_s G_q(x,x^2)G_q(y,y^2)+ P_t xy G_r(x,x^2)G_r(y,y^2),
\ee
and $S_0(x) [=B_0(x,1)=B_0(1,x)]$ is the generating function for $Q_0(k)$,
\be
S_0(x)= \sum_{k=0}^\infty Q_0(k) x^{k} = P_s G_q(x,x^2)+ P_t x G_r(x,x^2).
\ee
We note that the RCN model reduces to the configuration model when no triangle stubs exist, i.e., $p_{s,t}=p_s$, $q_{s,t}=(s+1)p_{s+1}/\langle s \rangle$, and $r_{s,t}=0$.
The generating functions are then written as $G_p(x,y)=G_{\rm tot}(x)=G_0(x)$, $G_q(x,y)=G_1(x)$, $G_r(x,y)=0$, $B_0(x,y)=G_1(x)G_1(y)$, and $S_0(x)=G_1(x)$, where $G_0(x)=\sum_{k} p_k x^k$ and $G_1(x)=G_0'(x)/G_0'(1)$.
It is easily confirmed from Eqs.~(\ref{eq:clusteringCoefficient}) and~(\ref{eq:r0}) that the configuration model is unclustered and uncorrelated; i.e., $C_0=0$ and $r_0=0$ \footnote{The configuration model can be highly clustered if the degree distribution is heavy tailed \cite{newman2018networks}. 
In the present study, we did not consider such a case.}.

\begin{figure}[b]
\begin{center}
\includegraphics[width=.35\textwidth]{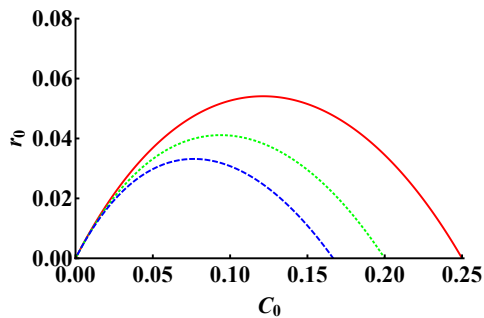}
\caption{
Assortativity $r_0$ of the Poisson RCN as a function of the clustering coefficient $C_0$. 
The red-solid, green-dotted, and blue-dashed lines represent $r_0$ for the cases of $\langle k \rangle=3$, $4$, and $5$, respectively.
}
\label{fig:assortativityR0}
\end{center}
\end{figure}

Let us apply the above-mentioned formulations to two types of RCNs.
The first example is the Poisson RCN, which has a double Poisson distribution 
\be
p_{s,t} = e^{-\langle s \rangle} \frac{\langle s \rangle^s}{s!} e^{-\langle t \rangle} \frac{\langle t \rangle^t}{t!}. \label{eq:doublePoisson}
\ee
In this case, $p_{s,t}=q_{s,t}=r_{s,t}$ and the generating functions are simplified to  $G_p(x,y)=G_q(x,y)=G_r(x,y)=e^{\langle s \rangle (x-1)}e^{\langle t \rangle (y-1)}$.
The clustering coefficient and the assortative coefficient are obtained from these generating functions as
\be
C_0=\frac{2 \langle t \rangle}{2\langle t \rangle+(\langle s \rangle+2\langle t \rangle)^2}
=\frac{2 \langle t \rangle}{2\langle t \rangle+\langle k \rangle^2},
\ee
and 
\be
r_0=\frac{2\langle s \rangle \langle t \rangle}{(\langle s \rangle + 2 \langle t \rangle)^3 + 2 \langle t \rangle (\langle s \rangle + 2 \langle t \rangle)^2 + 2 \langle s \rangle \langle t \rangle}=\frac{C_0 (1 - C_0(1 + \langle k \rangle))}{1 - C_0 (1+ (2 C_0-1) \langle k \rangle)}
\ge 0,
\ee
respectively. 
Figure~\ref{fig:assortativityR0} plots $r_0$ as a function of $C_0$ for several values of $\langle k \rangle$, showing that the Poisson RCN has a weak assortativity in the sense that $r_0$ takes a very small positive value when $0<C_0<1/(\langle k \rangle+1)$. 
It has been pointed out in \cite{miller2009percolation,gleeson2010clustering} that nodes assigned many triangles in the RCN possibly have high degrees compared to those assigned few triangles, although the edge and triangle stubs are randomly connected to stubs of the same type; this bias causes a positive correlation of the nearest degrees.

Another example is the delta RCN, which has a double $\delta$ function,
\be
p_{s,t} = \delta_{s,s_0}\delta_{t,t_0},
\ee
indicating that all nodes have $s_0$ single edges and $t_0$ triangles.
One immediately finds that $G_p(x,y)=x^{s_0}y^{t_0}$, $G_q(x,y)=x^{s_0-1}y^{t_0}$, and $G_r(x,y)=x^{s_0}y^{t_0-1}$; thus, $C_0=2t_0/(s_0+2t_0)(s_0+2t_0-1)$ and $r_0=0$. The delta RCN is clustered for $t_0>0$; however, it has no degree correlation because all nodes have the same degree $s_0+2t_0$.

\section{Structure of the percolating cluster in site percolation \label{sec:analysis}}

We consider site percolation on the RCNs: each node is occupied with probability $f$ and is unoccupied (removed from the original network) otherwise.
For site percolation on a network, the PC emerges at $f=f_c$, which is called the percolation threshold. 
The fraction $S$ of nodes belonging to the PC becomes $S>0$ ($S=0$) when $f> f_{\rm c}$ ($f \le f_c$).

The analytical treatment for site percolation on the RCN is presented below. 
We denote by $u$ the probability that a node reached by traversing a single edge chosen randomly from the original network is not a member of the PC and by $v$ the probability that a node reached by traversing a triangle edge is not a member of the PC.
Probabilities $u$ and $v$ are given as the solution of the following self-consistent equations, 
\be
u=\tf + f G_q(u,v^2) \label{eq:recUsite}
\ee
and
\be
v^2=(\tf+f G_r(u,v^2))^2, \label{eq:recVsite}
\ee
where $\tf=1-f$. 
The normalized PC size, $S$, is given as the probability that a randomly chosen node is a member of the PC; hence, we have the following equation:
\be
S = f(1-\sum_{s,t} p_{s,t} u^s v^{2t}) = f(1-G_p(u,v^2)).
\ee

To find the percolation threshold $f_c$, we assume $u=1-\epsilon_u$ and $v^2=1-\epsilon_v$ and consider the stability of the trivial solution $(u,v)=(1,1)$. 
Expanding Eqs.~(\ref{eq:recUsite}) and (\ref{eq:recVsite}) to leading order in $\epsilon_u$ and $\epsilon_v$ gives ${\bm \epsilon} = A {\bm \epsilon}$, where ${\bm \epsilon}=[\epsilon_u, \epsilon_v]^{\mathrm T}$ and
\be
A=
f \left[
\begin{array}{cc}
\sum s q_{s,t} & \sum t q_{s,t} \\
2 \sum s r_{s,t} & 2 \sum t r_{s,t}
\end{array}
\right].
\label{eq:matrixSite}
\ee
The percolation threshold $f_c$ is then given from the condition $\det|A-I|=0$ ($I$ is the identity matrix).
The following two subsections focus on the PC for $f>f_c$, where Eqs.~(\ref{eq:recUsite}) and (\ref{eq:recVsite}) have a nontrivial solution of $u$ and $v$, i.e., $S>0$, deriving its clustering coefficient and assortative coefficient. 

\subsection{Clustering coefficient of the percolating cluster}\label{sec:cluster}

We derive the clustering coefficient of the PC by starting with the conditional probability $P({\rm PC}, s, t_1, t_2 | m, n)$ that a randomly chosen node belongs to the PC and has $s$ single edges, $t_1$ triangles with one removed, and $t_2$ triangles (i.e., the node has $t_1+2t_2$ triangle edges) in the PC, given that it has $m$ single edges and $n$ triangles in the original network. This probability is given as 
\be
P({\rm PC}, s, t_1, t_2 | m, n) = f \binom{m}{s} f^{s} \tf^{m-s} \binom{n}{t_2} f^{2t_2} \binom{n-t_2}{t_1} (2f \tf)^{t_1} \tf^{2(n-t_2-t_1)}\Big(1-\tu^{s}\tv^{2t_2+t_1}\Big),
\ee
where 
\be
\tu=G_q(u,v^2) \quad {\rm and} \quad \tv=G_r(u,v^2)
\ee
are the probability that the occupied node reached by traversing a single edge is not a member of the PC and the probability that the occupied node reached by traversing a triangle edge is not a member of the PC, respectively.
Because the probability $P({\rm PC}, s, t_1, t_2)$ that a randomly chosen node belongs to the PC and has $s$ single edges, $t_1$ triangles with one removed, and $t_2$ triangles in the PC is
\begin{eqnarray}
P({\rm PC}, s, t_1, t_2)
&=& 
\sum_{m,n} P({\rm PC}, s, t_1, t_2 | m, n) p_{m,n} \nonumber \\
&=& 
f \sum_{m,n}p_{m,n}\binom{m}{s} f^{s} \tf^{m-s} \binom{n}{t_2} f^{2t_2} \binom{n-t_2}{t_1} (2f \tf)^{t_1} \tf^{2(n-t_2-t_1)}\Big(1-\tu^{s}\tv^{2t_2+t_1}\Big),
\end{eqnarray}
and the probability $P({\rm PC})$ that a randomly chosen node belongs to the PC is
\be
P({\rm PC})=\sum_{s,t_1,t_2}P({\rm PC}, s, t_1, t_2)=f(1-G_p(u,v^2))=S,
\ee
we easily obtain the probability $P_{\rm PC}(s,t_1,t_2) \equiv P(s, t_1, t_2 |{\rm PC})$ that a randomly chosen node has $s$ single edges, $t_1$ triangles with one removed, and $t_2$ triangles conditioned on the node belonging to the PC, from $P_{\rm PC}(s,t_1,t_2)=P({\rm PC}, s, t_1, t_2)/P({\rm PC})$. 

Introducing the generating function $F_{\rm PC}(x,y,z)$ for $P_{\rm PC}(s,t_1,t_2)$ as 
\begin{eqnarray}
F_{\rm PC}(x,y,z) 
&=& \sum_{s,t_1,t_2} P_{\rm PC}(s,t_1,t_2) x^{s}y^{t_1}z^{t_2} \\
&=& \frac{1}{1-G_p(u,v^2)} \Big( G_p(f x+\tf,f^2 z+2f \tf y + \tf^2)-G_p(f \tu x+\tf,f^2 \tv^2 z+2f \tf \tv y + \tf^2) \Big), \nonumber
\end{eqnarray}
we obtain the degree distribution $P_{\rm PC}(k)$ of the PC and the clustering coefficient $C_{\rm PC}$ of the PC as follows:
\be
P_{\rm PC}(k) = \frac{1}{k!} \frac{\partial^k}{\partial x^k} F_{\rm PC}(x,x,x^2) \Big|_{x=0},
\ee
and
\be
C_{\rm PC}=\frac{\partial}{\partial z} F_{\rm PC}(x,y,z) \Big|_{x=y=z=1}\Big/ \frac{1}{2} \frac{\partial^2}{\partial x^2} F_{\rm PC}(x,x,x^2) \Big|_{x=1}.
\ee
We note that for the case of $p_{s,t}=p_s$, $F_{\rm PC}(x,y,z)$ is independent of $y$ and $z$, so $C_{\rm PC}=0$.
It means that the PC formed by site percolation on the configuration model is unclustered.

\subsection{Assortative coefficient of the percolating cluster}\label{sec:assortativity}

Next, we formalize the assortative coefficient of the PC.
Our derivation is an extension of \cite{mizutaka2018disassortativity} in which the assortative coefficient of the PC formed by site percolation on  uncorrelated networks was derived. 

First, we consider the conditional probability $Q_s({\rm PC}, s_1, t_1, s_2, t_2 | m_1, n_1, m_2, n_2)$ that a single edge belongs to the PC and its one end has $s_1$ other single edges and $t_1$ triangle edges and the other end has $s_2$ other single edges and $t_2$ triangle edges in the PC, given that the two ends of the selected single edge have $m_1$ other single edges and $n_1$ triangles and $m_2$ other single edges and $n_2$ triangles in the original network, respectively.
This probability is written as follows:
\begin{eqnarray}
&&Q_s({\rm PC}, s_1, t_1, s_2, t_2 | m_1, n_1, m_2, n_2) \\
&&= 
f^2 \binom{m_1}{s_1} f^{s_1} \tf^{m_1-s_1} \binom{2n_1}{t_1} f^{t_1} \tf^{2n_1-t_1} 
\binom{m_2}{s_2} f^{s_2} \tf^{m_2-s_2} \binom{2n_2}{t_2} f^{t_2} \tf^{2n_2-t_2} 
(1-\tu^{s_1+s_2}\tv^{t_1+t_2}).
\nonumber
\end{eqnarray}
Here, $f^2$ in the right-hand side represents the probability that two ends of the focal edge are not removed and thus the edge remains.
The probability $Q_s({\rm PC}, s_1, t_1, s_2, t_2)$ that a single edge belongs to the PC and its one end has $s_1$ other single edges and $t_1$ triangle edges and the other end has $s_2$ other single edges and $t_2$ triangle edges in the PC is
\be
Q_s({\rm PC}, s_1, t_1, s_2, t_2)=\sum_{m_1 \ge s_1}\sum_{m_2 \ge s_2}\sum_{2 n_1 \ge t_1}\sum_{2 n_2 \ge t_2} Q_s(m_1,n_1,m_2,n_2) Q_s({\rm PC}, s_1, t_1, s_2, t_2 | m_1, n_1, m_2, n_2).
\ee
Here, $Q_s(m_1,n_1,m_2,n_2)$ is the probability that a single edge has two ends: one has $m_1$ other single edges and $n_1$ triangles and the other has $m_2$ other single edges and $n_2$ triangles in the original network. This probability is written as $Q_s(m_1,n_1,m_2,n_2)=q_{m_1,n_1}q_{m_2,n_2}$ in that the RCN is a random network.
The corresponding generating function for $Q_s({\rm PC}, s_1, t_1, s_2, t_2)$ is as follows:
\begin{eqnarray}
H_s(x_1,x_2,y_1,y_2) 
&=& \sum_{s_1=0}^\infty \sum_{t_1=0}^\infty \sum_{s_2=0}^\infty \sum_{t_2=0}^\infty Q_s({\rm PC}, s_1, t_1, s_2, t_2) x_1^{s_1}x_2^{t_1}y_1^{s_2}y_2^{t_2}
\nonumber \\
&=& f^2 \Big(\mgq(x_1,x_2)\mgq(y_1,y_2)-\mgq(\tu x_1,\tv x_2)\mgq(\tu y_1,\tv y_2) \Big),
\end{eqnarray}
where
\be
\mgq(x,y)=G_q(\tf+fx,(\tf+fy)^2).
\ee

We further introduce the conditional probability $Q_t({\rm PC}, s_1, t_1, s_2, t_2 | m_1, n_1, m_2, n_2)$ that a triangle edge belongs to the PC and one end of the edge has $s_1$ single edges and $t_1$ other triangle edges and the other end has $s_2$ single edges and $t_2$ other triangle edges in the PC, respectively, given that the two ends of the selected triangle edge have $m_1$ single edges and $n_1$ other triangles (triangles except the one including the selected edge) and $m_2$ single edges and $n_2$ other triangles in the original network, respectively, as 
\begin{eqnarray}
&&Q_t({\rm PC}, s_1, t_1, s_2, t_2 | m_1, n_1, m_2, n_2) \\
&&=
f^2 \tf \binom{m_1}{s_1} f^{s_1} \tf^{m_1-s_1} \binom{2n_1}{t_1} f^{t_1} \tf^{2n_1-t_1} 
\binom{m_2}{s_2} f^{s_2} \tf^{m_2-s_2} \binom{2n_2}{t_2} f^{t_2} \tf^{2n_2-t_2} 
(1-\tu^{s_1+s_2}\tv^{t_1+t_2}) 
\nonumber \\
&&\;\;\;+f^3 \binom{m_1}{s_1} f^{s_1} \tf^{m_1-s_1} \binom{2n_1}{t_1-1} f^{t_1-1} \tf^{2n_1-t_1+1} 
\binom{m_2}{s_2} f^{s_2} \tf^{m_2-s_2} \binom{2n_2}{t_2-1} f^{t_2-1} \tf^{2n_2-t_2+1} 
(1-\tu^{s_1+s_2}\tv^{t_1+t_2-1}).
\nonumber
\end{eqnarray}
The two ends of a triangle edge have a common neighbor to form a triangle. 
The first and the second terms of the right-hand side are the contributions when this neighbor is unoccupied and occupied, respectively.
The probability $Q_t({\rm PC}, s_1, t_1, s_2, t_2)$ that a triangle edge belongs to the PC and its ends have $s_1$ single edges and $t_1$ other triangle edges and $s_2$ single edges and $t_2$ other triangle edges in the PC, respectively, is
\begin{eqnarray}
Q_t({\rm PC}, s_1, t_1, s_2, t_2)
&=&\sum_{m_1 \ge s_1}\sum_{m_2 \ge s_2}\sum_{2 n_1 \ge t_1}\sum_{2 n_2 \ge t_2} Q_t(m_1,n_1, m_2,n_2) Q_t({\rm PC}, s_1, t_1, s_2, t_2 | m_1, n_1, m_2, n_2), 
\end{eqnarray}
where the probability $Q_t(m_1,n_1, m_2,n_2)$ that the ends of a triangle edge have $m_1$ single edges and $n_1$ other triangles and $m_2$ single edges and $n_2$ other triangles in the original network, respectively, is $Q_t(m_1,n_1, m_2,n_2)=r_{m_1,n_1}r_{m_2,n_2}$ for the RCN. 
The corresponding generating function for $Q_t({\rm PC}, s_1, t_1, s_2, t_2)$ is as follows:
\begin{eqnarray}
H_t(x_1,x_2,y_1,y_2) 
&=& \sum_{s_1=0}^\infty \sum_{t_1=0}^\infty \sum_{s_2=0}^\infty \sum_{t_2=0}^\infty Q_t({\rm PC}, s_1, t_1, s_2, t_2) x_1^{s_1}x_2^{t_1}y_1^{s_2}y_2^{t_2}
\nonumber \\
&=& 
f^2 \Big( (\tf + f x_2 y_2) \mgr(x_1,x_2)\mgr(y_1,y_2) - (\tf + f \tv x_2 y_2) \mgr(\tu x_1,\tv x_2)\mgr(\tu y_1, \tv y_2) \Big),
\end{eqnarray}
where
\be
\mgr(x,y)=G_r(\tf+fx, (\tf+fy)^2).
\ee

The probability $Q({\rm PC}, s_1, t_1, s_2, t_2)$ that an edge belongs to the PC and the ends of the selected edge have $s_1$ single edges and $t_1$ triangle edges and $s_2$ single edges and $t_2$ triangle edges except the selected edge in the PC, respectively, is
\begin{eqnarray}
Q({\rm PC}, s_1, t_1, s_2, t_2) = P_s Q_s({\rm PC}, s_1, t_1, s_2, t_2) +P_t Q_t({\rm PC}, s_1, t_1, s_2, t_2),
\end{eqnarray}
in that an edge chosen randomly from the original network is either of a single edge (with probability $P_s$) or a triangle edge (with probability $P_t$).
The corresponding generating function is given as follows:
\begin{eqnarray}
H(x_1,x_2,y_1,y_2)
&&= \sum_{s_1=0}^\infty \sum_{t_1=0}^\infty \sum_{s_2=0}^\infty \sum_{t_2=0}^\infty Q({\rm PC}, s_1, t_1, s_2, t_2) x_1^{s_1}x_2^{t_1}y_1^{s_2}y_2^{t_2}
\nonumber \\
&&= P_s H_s(x_1,x_2,y_1,y_2)+ P_t H_t(x_1,x_2,y_1,y_2).
\end{eqnarray}
The corresponding generating function for the probability $Q({\rm PC}, k_1, k_2)$ that an edge belongs to the PC and has two ends with degrees $k_1+1$ and $k_2+1$ in the PC is given as $H(x,x,y,y)$.
The probability $Q({\rm PC})$ that a randomly chosen edge belongs to a PC is given as follows:
\begin{eqnarray}
Q({\rm PC}) 
&=& H(1,1,1,1) \nonumber \\
\nonumber \\
&=& f^2 (P_s (1-\tu^2) + P_t(1-(\tf+f \tv)\tv^2)).
\end{eqnarray}
Hence, we obtain the probability $Q_{\rm PC}(k_1, k_2) \equiv Q(k_1, k_2 | {\rm PC})$ that an edge chosen from the PC has two ends with degree $k_1+1$ and $k_2+1$ in the PC from $Q_{\rm PC}(k_1, k_2) = Q({\rm PC}, k_1, k_2)/Q({\rm PC})$.

Using the generating function $B_{\rm PC}(x,y)$ for $Q_{\rm PC}(k_1, k_2)$ given as
\begin{eqnarray}
B_{\rm PC}(x,y)
&=&\sum_{k_1 \ge 0}\sum_{k_2 \ge 0} Q_{\rm PC}(k_1, k_2) x^{k_1} y^{k_2} =\frac{H(x,x,y,y)}{Q({\rm PC})}  \nonumber \\
&=& 
\frac{f^2P_s}{Q({\rm PC})} \Big[ \mgq(x,x)\mgq(y,y)-\mgq(\tu x, \tv x)\mgq(\tu y, \tv y)\Big] \nonumber \\
&&+\frac{f^2 P_t}{Q({\rm PC})} \Big[(\tf+f x y)\mgr(x,x)\mgr(y,y) - (\tf+f \tv x y)\mgr(\tu x,\tv x)\mgr(\tu y,\tv y) \Big],
\end{eqnarray}
and the generating function $S_{\rm PC}(x)[=B_{\rm PC}(x,1)=B_{\rm PC}(1,x)]$ for the probability $Q_{\rm PC}(k)=\sum_{k'} Q_{\rm PC}(k, k')$ of an edge in the PC reaching a node with degree $k+1$, given as
\begin{eqnarray}
S_{\rm PC}(x) &=&\sum_{k \ge 0} Q_{\rm PC}(k) x^{k} 
= \frac{H(x,x,1,1)}{Q({\rm PC})}  \nonumber \\ 
&=& 
\frac{f^2P_s}{Q({\rm PC})} \Big[ \mgq(x,x) - \tu \mgq(\tu x, \tv x)\Big] 
\nonumber \\
&&
+\frac{f^2 P_t}{Q({\rm PC})} \Big[(\tf+f x)\mgr(x,x) - (\tf+f \tv x) \tv \mgr(\tu x,\tv x) \Big],
\end{eqnarray} 
we obtain the assortative coefficient $r_{\rm PC}$ of the PC as 
\be
r_{\rm PC}=\frac{\partial_x \partial_y B_{\rm PC}(x,y) - (\partial_x S_{\rm PC}(x))^2}{(x \partial_x)^2 S_{\rm PC}(x) - (\partial_x S_{\rm PC}(x))^2} \Big|_{x=y=1}. \label{eq:rPCfinal}
\ee
This is an extension of the formulation for uncorrelated networks \cite{mizutaka2018disassortativity}. 
For the case of $p_{s,t}=p_s$, one finds, after tedious but simple algebra, that the above formulation yields the assortative coefficient of the PC on the uncorrelated networks:
$r_{\rm PC}$ is given by Eq.~(\ref{eq:rPCfinal}) with $B_{\rm PC}(x,y) = [G_1(\tilde{f}+fx)G_1(\tilde{f}+fy)-G_1(\tilde{f}+f \tu x)G_1(\tilde{f}+f \tu y)]/(1-\tu^2)$ and $S_{\rm PC}(x) = [G_1(\tilde{f}+fx)-\tu G_1(\tilde{f}+f \tu x)]/(1-\tu^2)$, where $\tu=G_1(u)$ and $u$ is the solution of $u=\tilde{f}+f G_1(u)$. 

\begin{figure}[!b]
\begin{center}
\includegraphics[height=.23\textwidth]{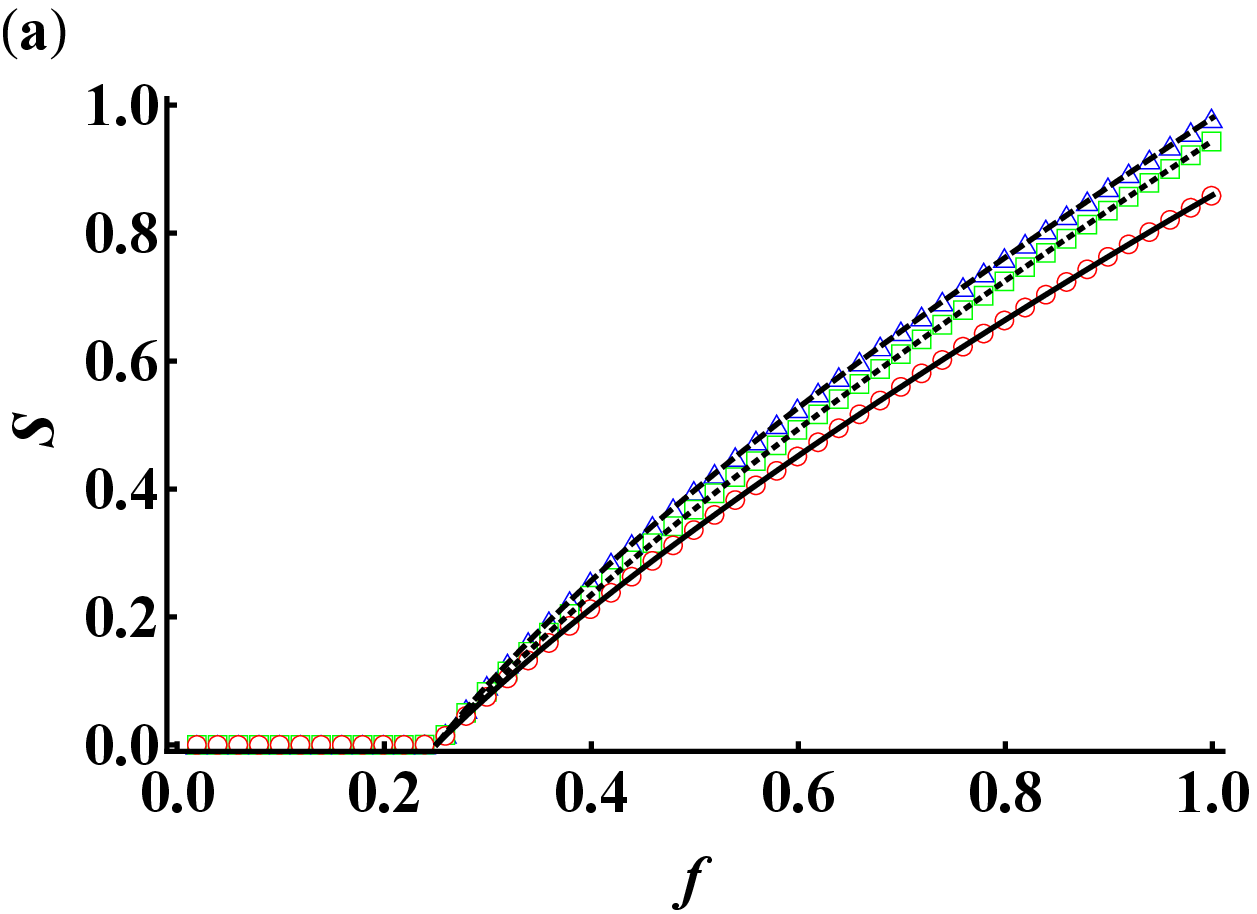}
\includegraphics[height=.23\textwidth]{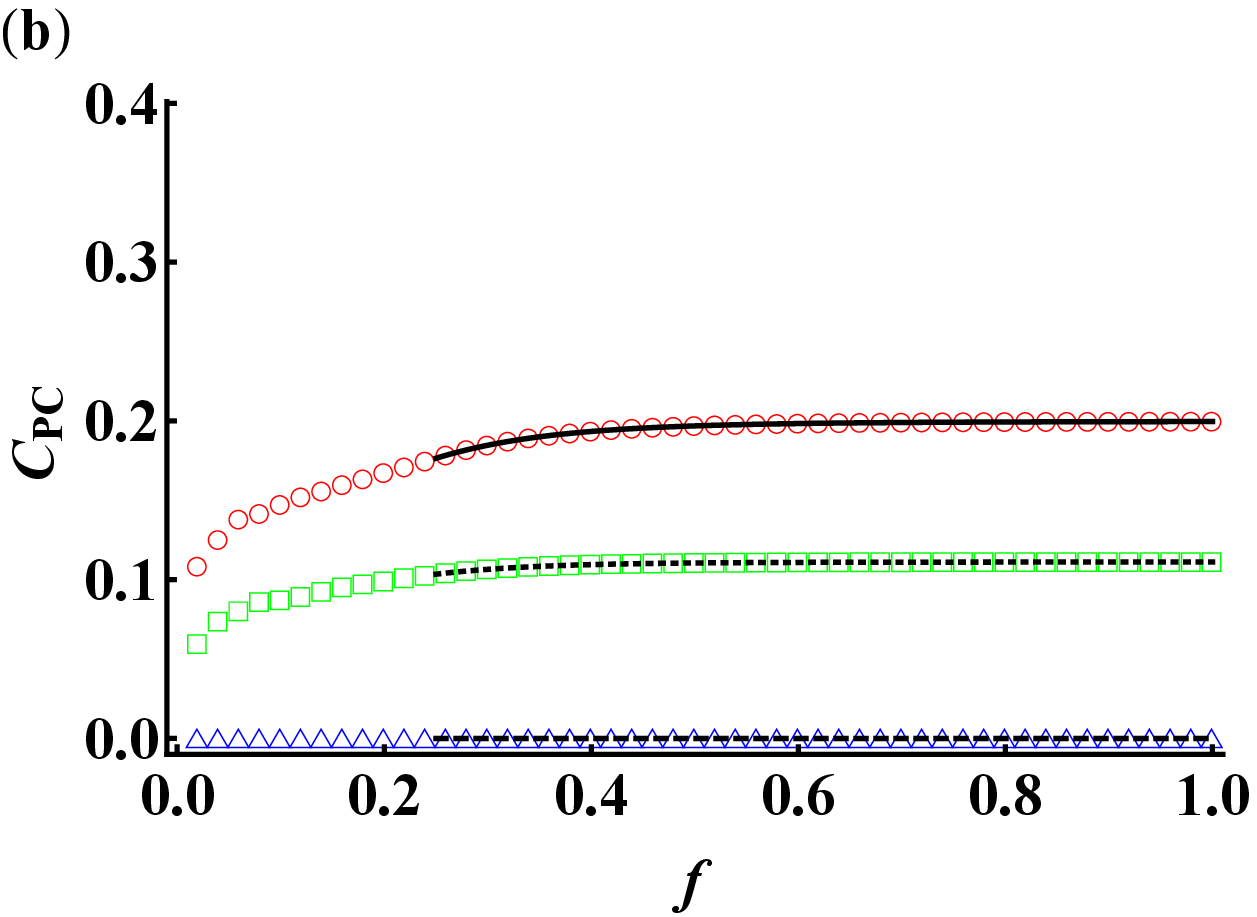}
\includegraphics[height=.23\textwidth]{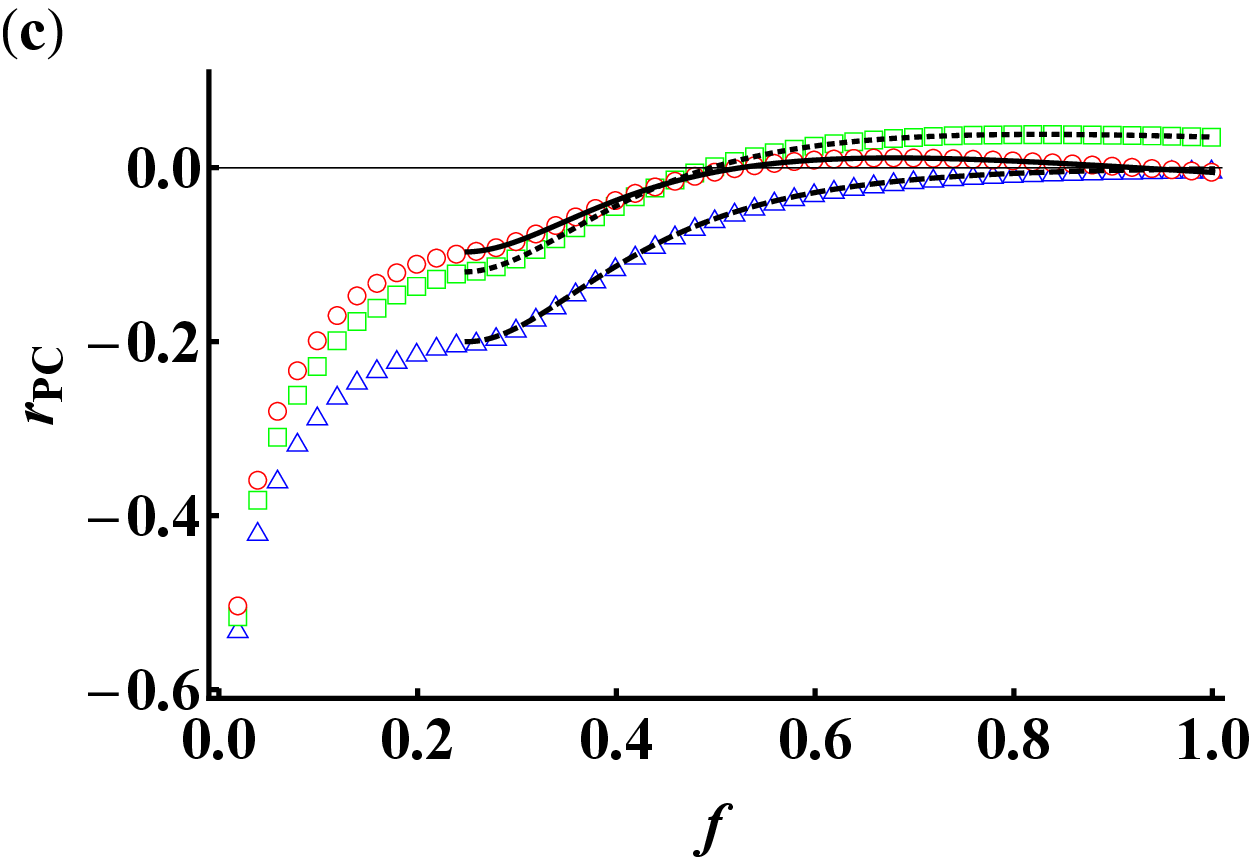}
\caption{
Comparison with the analytical treatments and simulation results for the structures of the PC formed by site percolation: (a) normalized size $S$, (b) clustering coefficient $C_{\rm PC}$, and (c) assortative coefficient $r_{\rm PC}$. 
The Poisson RCN with $\langle k \rangle=4$ is utilized as the original network. 
The simulation results are for the cases of $\langle s \rangle=0$ and $\langle t \rangle=2$ (red circles), $\langle s \rangle=2$ and $\langle t \rangle=1$ (green squares), and $\langle s \rangle=4$ and $\langle t \rangle=0$ (blue triangles).
The corresponding analytical estimates are represented by the solid lines, dotted lines, and dashed lines.
}
\label{fig:poissonRCN:site}
\end{center}
\end{figure}

\begin{figure}[!h]
\begin{center}
\includegraphics[height=.28\textwidth]{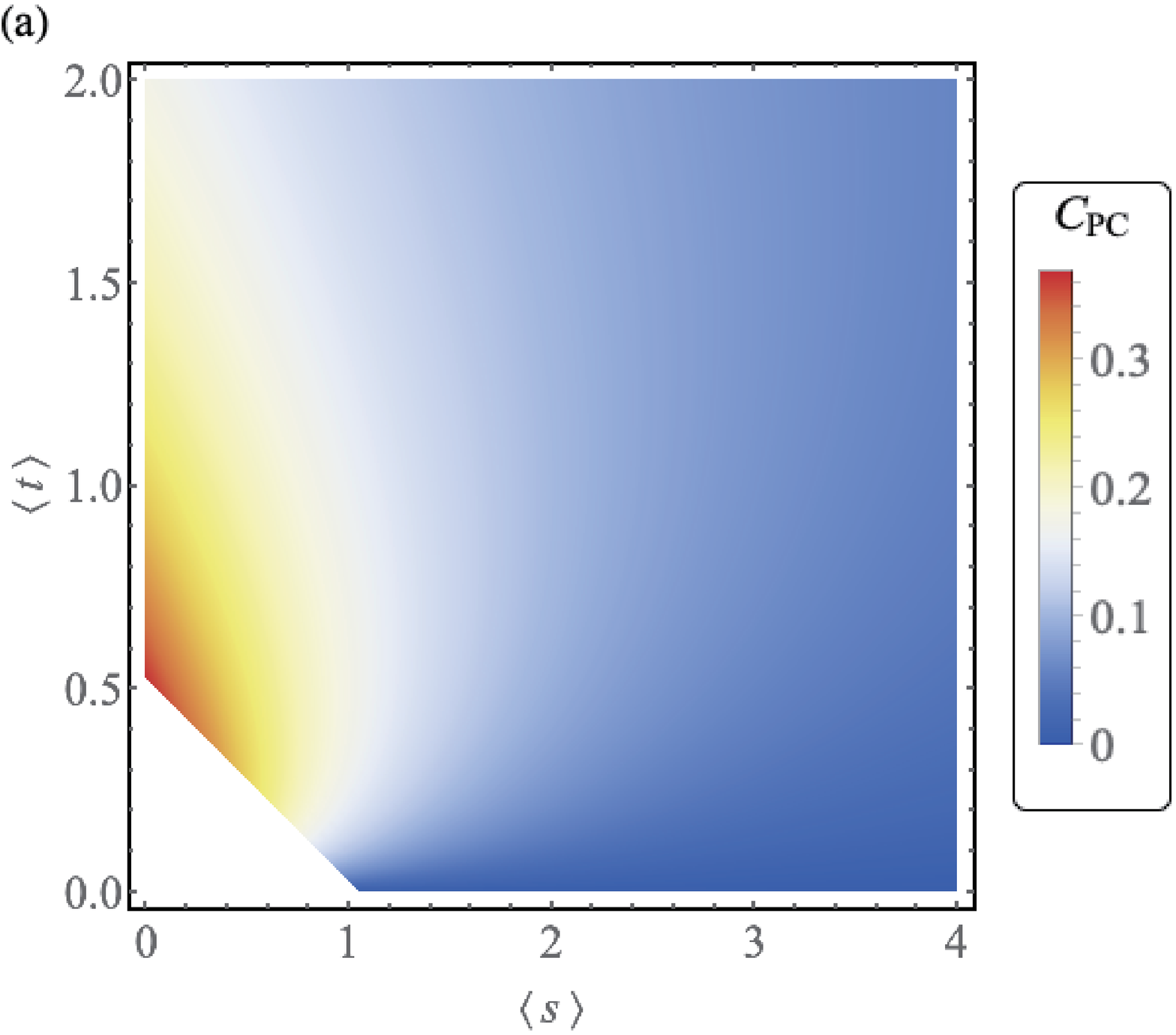}
\includegraphics[height=.28\textwidth]{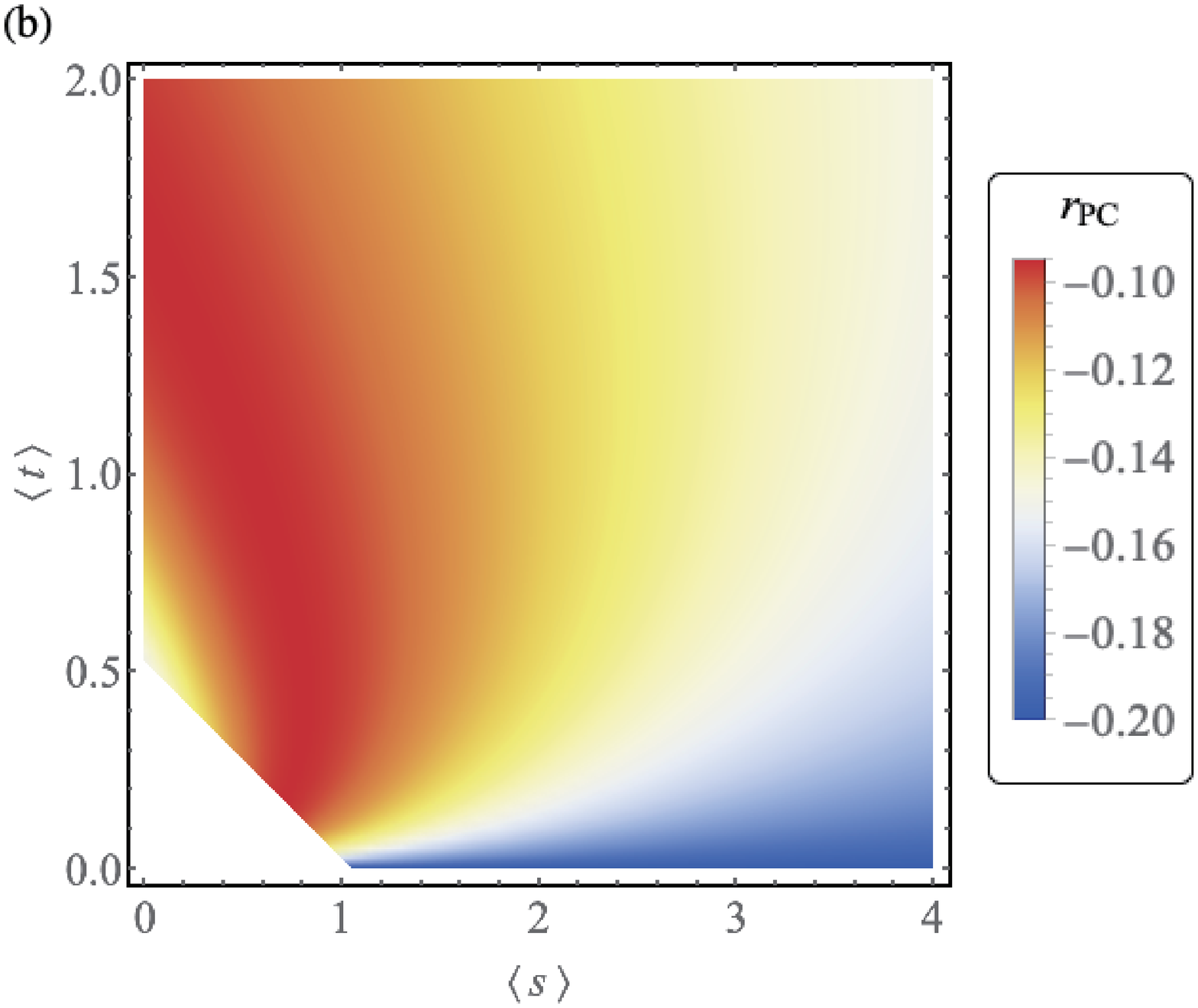}
\includegraphics[height=.22\textwidth]{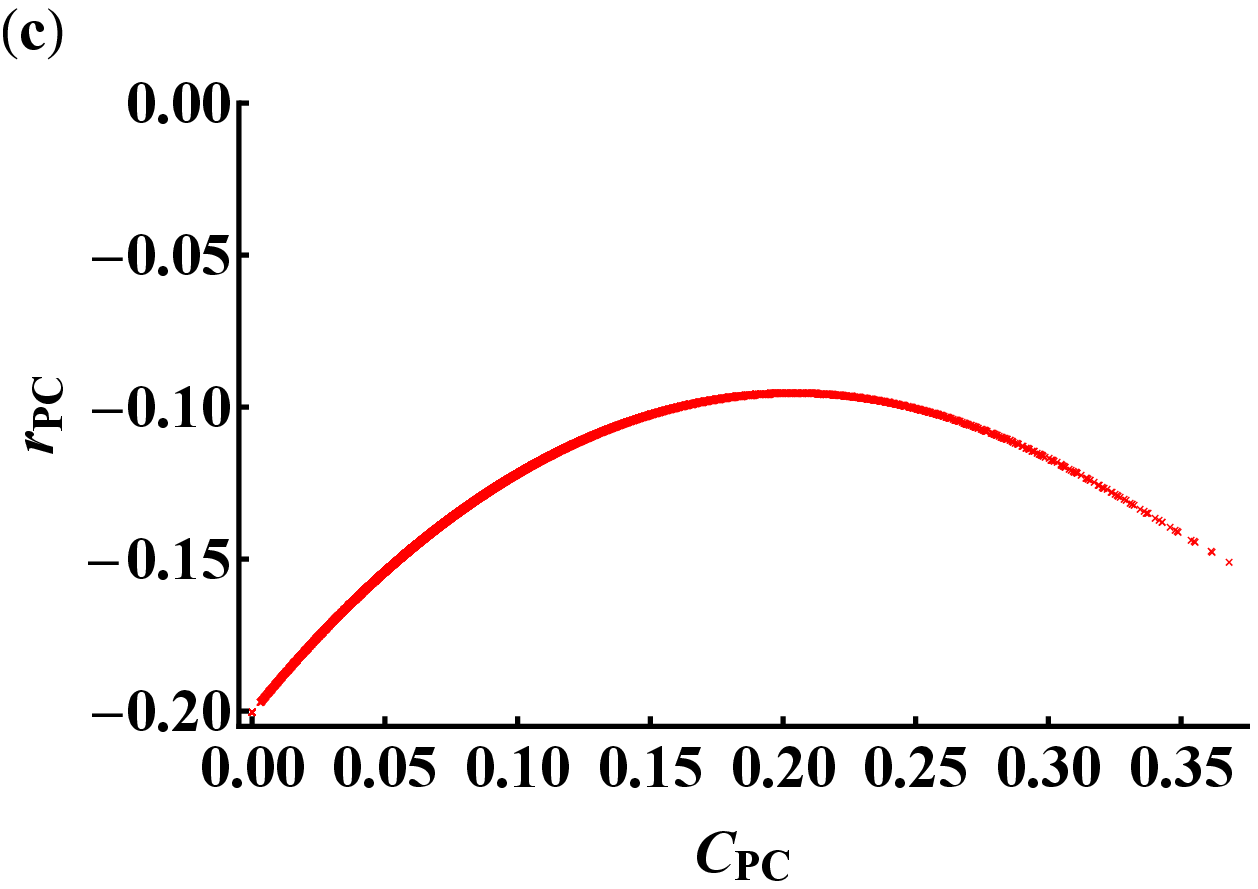}
\caption{
Dependence of (a) clustering coefficient $C_{\rm PC}$ and (b) assortative coefficient $r_{\rm PC}$ of the fractal PC formed by site percolation on $\langle s \rangle$ and $\langle t \rangle$ of the Poisson RCN. 
Here $C_{\rm PC}$ and $r_{\rm PC}$ at $f \simeq f_c$ are obtained from analytical estimates.
Blank areas reflect the absence of the PC.
(c) Scatter plot of $C_{\rm PC}$ (color-coded values in (a)) and $r_{\rm PC}$ (color-coded values in (b)) on the fractal PC.
}
\label{fig:poissonRCN:site:diagram}
\end{center}
\end{figure}

\begin{figure}[!h]
\begin{center}
\includegraphics[height=.23\textwidth]{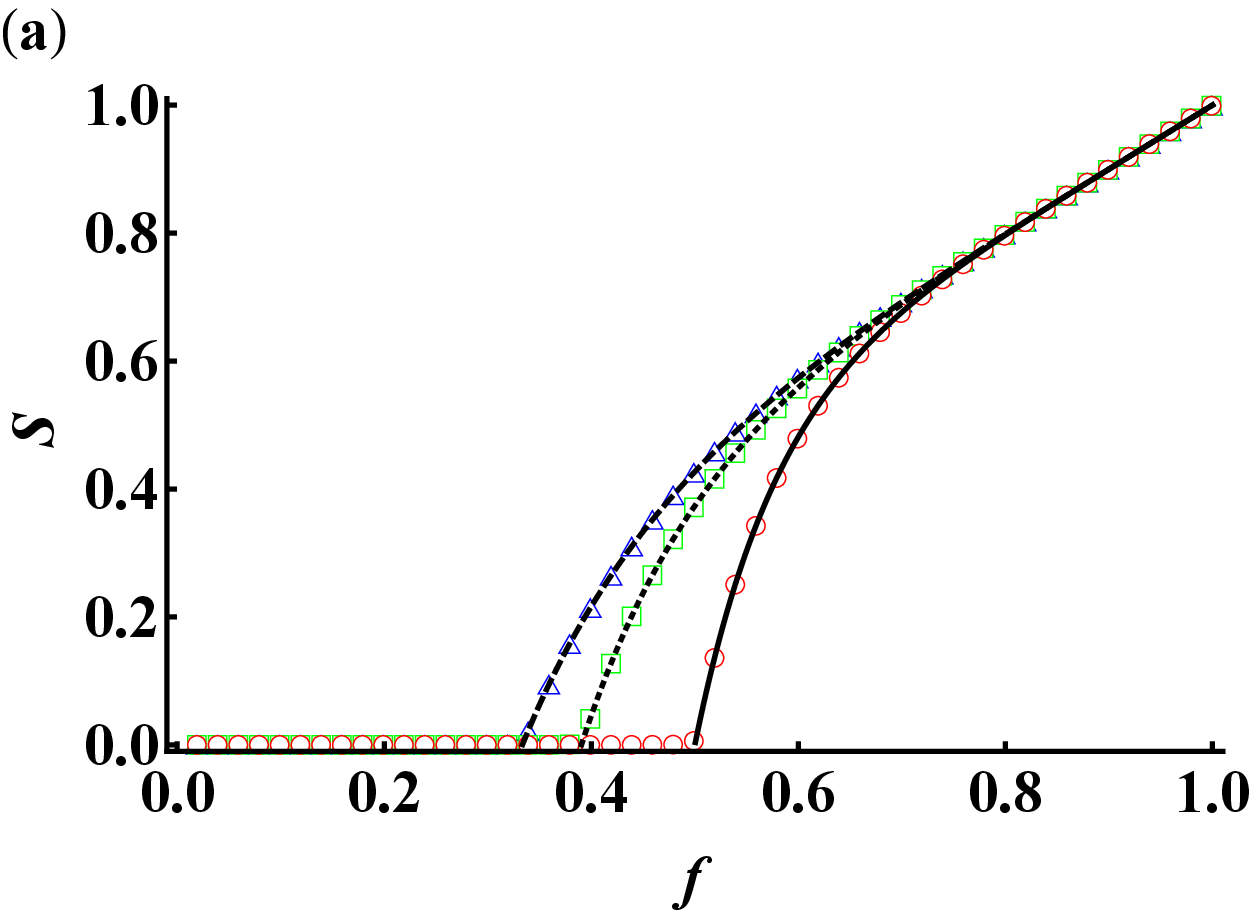}
\includegraphics[height=.23\textwidth]{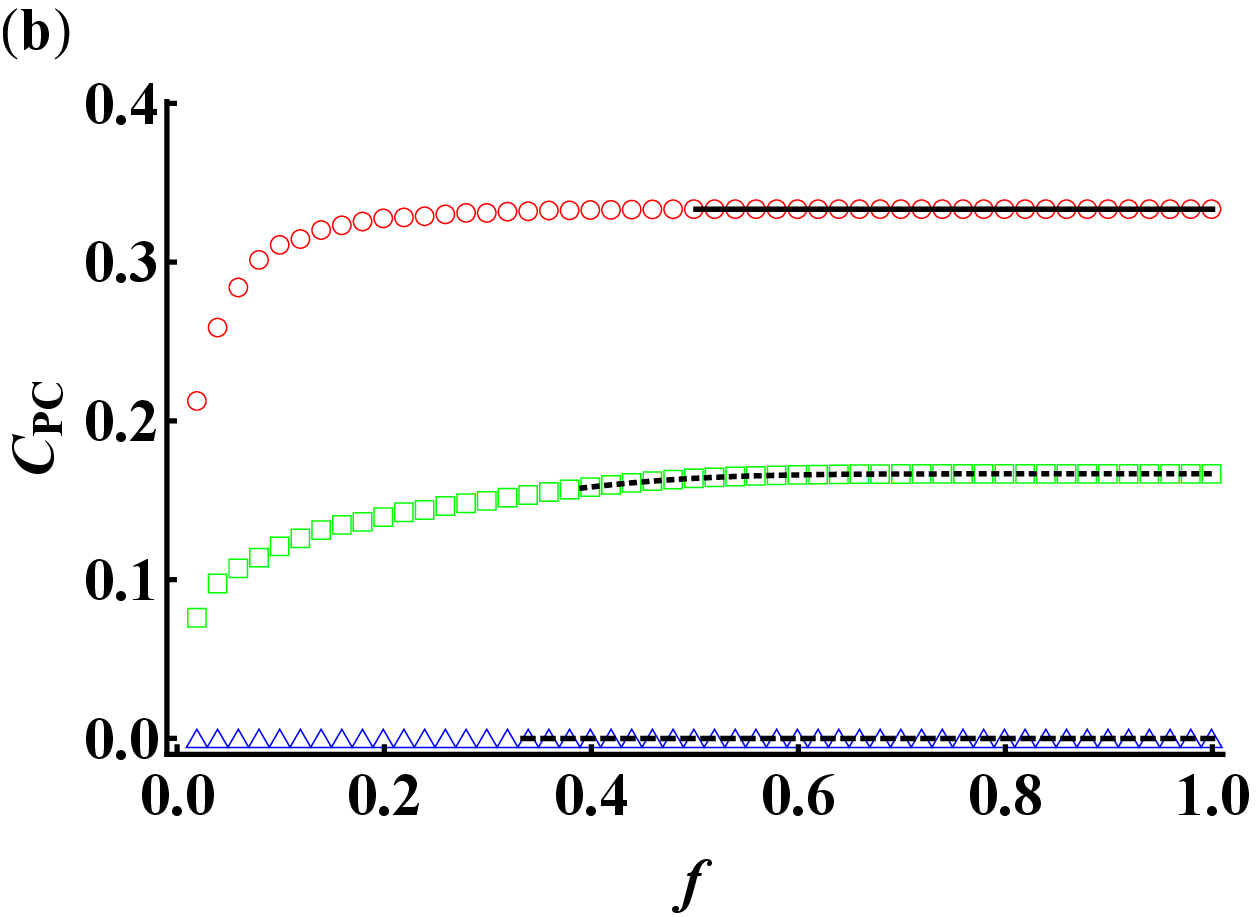}
\includegraphics[height=.23\textwidth]{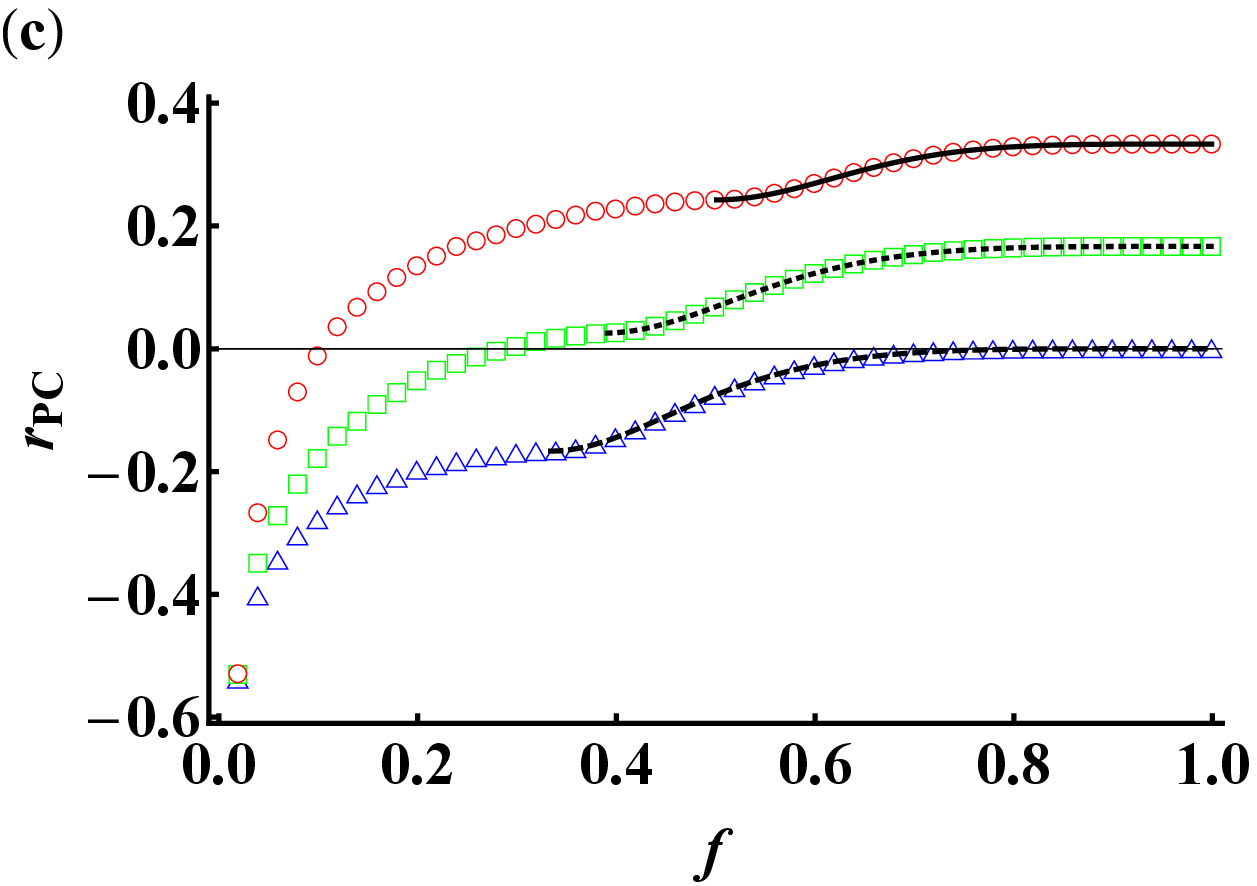}
\caption{
Comparison with the analytical treatments and simulation results for the structures of the PC formed by site percolation: (a) normalized size $S$, (b) clustering coefficient $C_{\rm PC}$, and (c) assortative coefficient $r_{\rm PC}$. 
The delta RCN with degree 4 is utilized as the original network. 
The simulation results are for the cases of $s_0=0$ and $t_0=2$ (red circles), $s_0=2$ and $t_0=1$ (green squares), and $s_0=4$ and $t_0=0$ (blue triangles).
The corresponding analytical estimates are represented by the solid lines, dotted lines, and dashed lines.
}
\label{fig:deltaRCN:site}
\end{center}
\end{figure}

\begin{figure}[!h]
\begin{center}
\includegraphics[height=.21\textwidth]{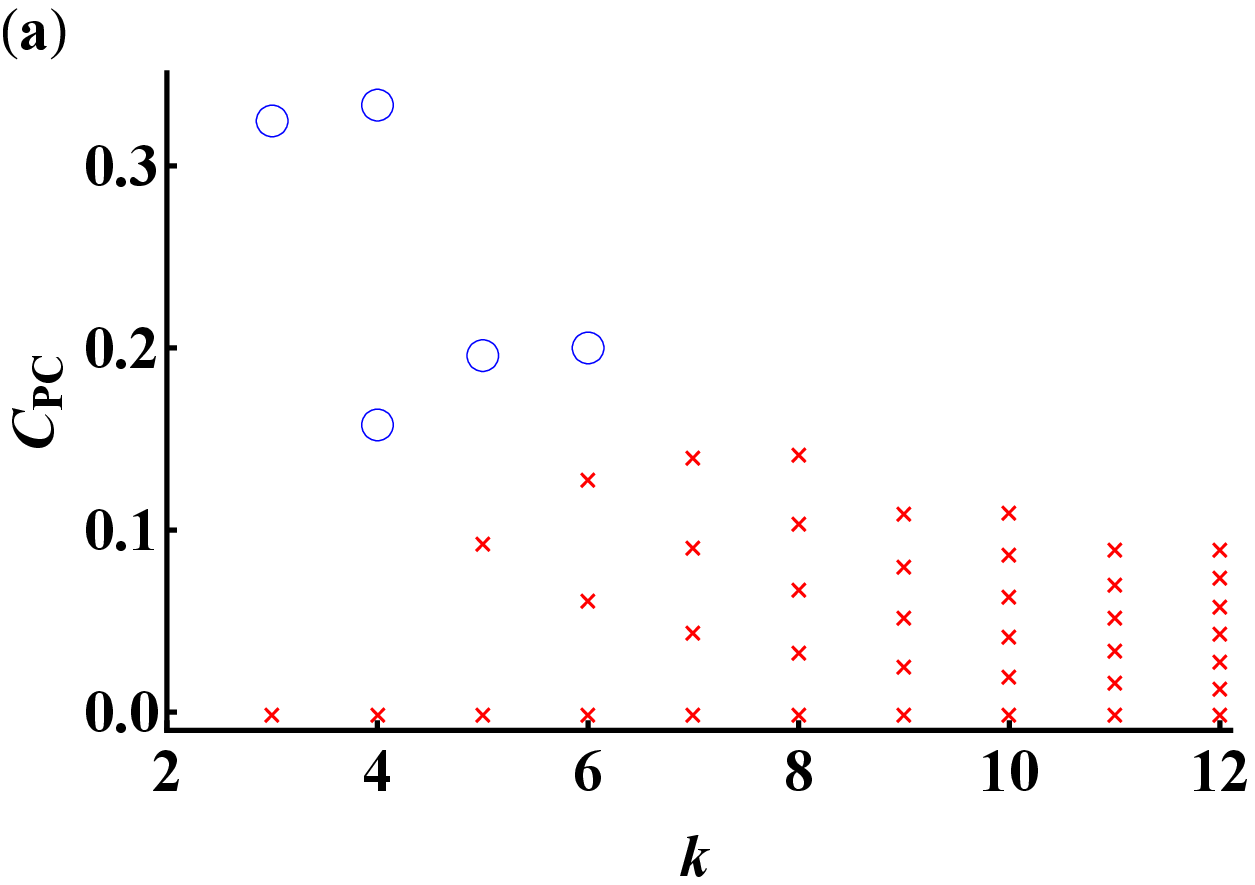}
\includegraphics[height=.21\textwidth]{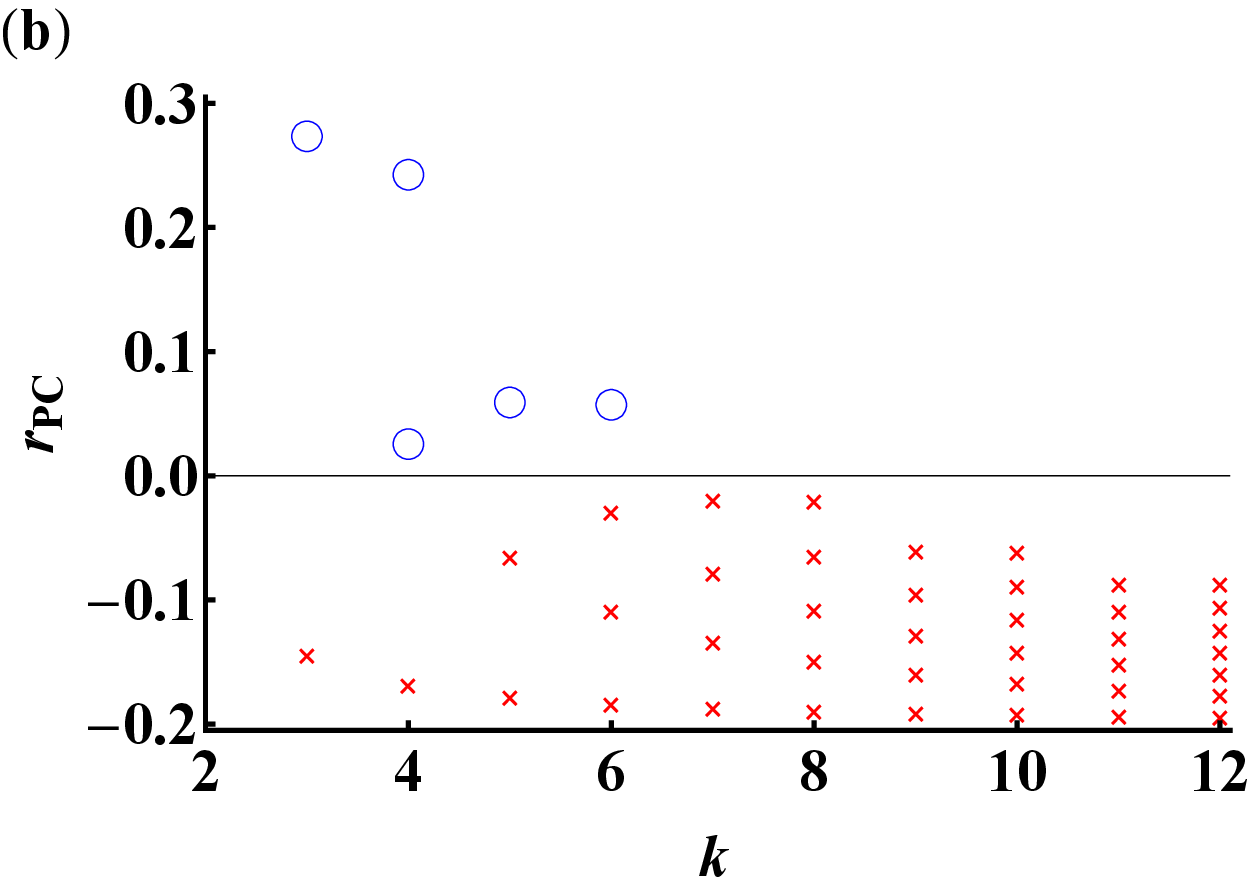}
\includegraphics[height=.21\textwidth]{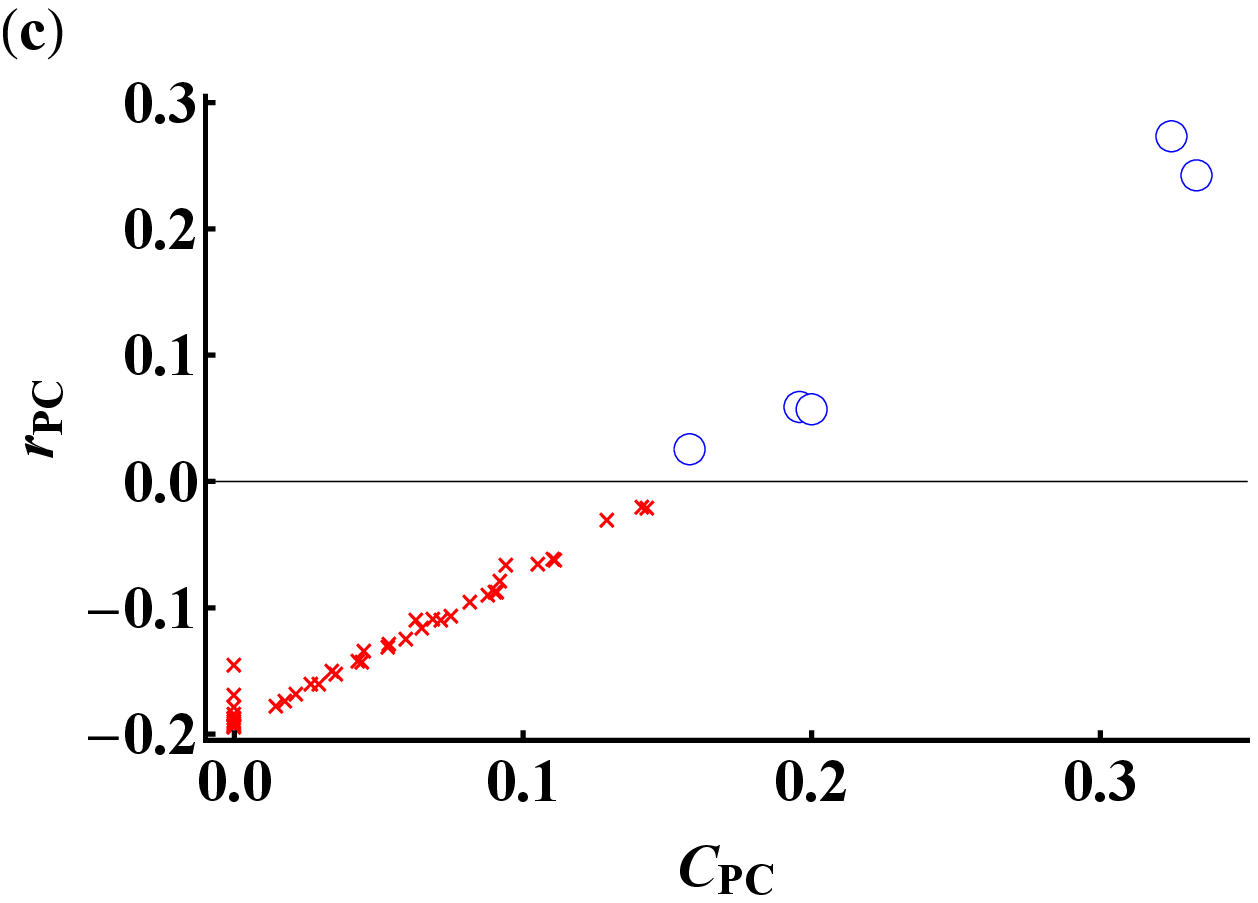}
\caption{
Result for (a) clustering coefficient $C_{\rm PC}$ and (b) assortative coefficient $r_{\rm PC}$ of the fractal PC formed by site percolation on the delta RCN with degree $k$ and (c) their scatter plot.
Here $C_{\rm PC}$ and $r_{\rm PC}$ at $f \simeq f_c$ are obtained from analytical estimates.
All possible combinations of $s_0$ and $t_0$ for given degree $k (=3, 4, \cdots, 12)$ are considered.
The blue circles (red crosses) represent data indicating $r_{\rm PC}>0$ ($r_{\rm PC} < 0$).
}
\label{fig:deltaRCN:site:diagram}
\end{center}
\end{figure}

\subsection{Examples with numerical check}\label{sec:numericalcheck}

In this subsection, we applied our analysis to two RCNs, namely, the Poisson RCN and the delta RCN, discussing the structural properties of the PC formed by site percolation.
Moreover, we performed Monte Carlo simulations to verify the validity of our analytical estimates.
In our simulations, we generated 10 network realizations consisting of $N=3 \times 10^6$ nodes and carried out the Newman-Ziff algorithm \cite{newman2001fast} for site percolation $10^3$ times on each realization. 
On each run, we specified the largest cluster corresponding to the PC for $f > f_{\rm c}$ and evaluated its size, clustering coefficient, and assortative coefficient to compare each average value with the corresponding analytical estimate.

\begin{figure}[!h]
\begin{center}
\includegraphics[height=.23\textwidth]{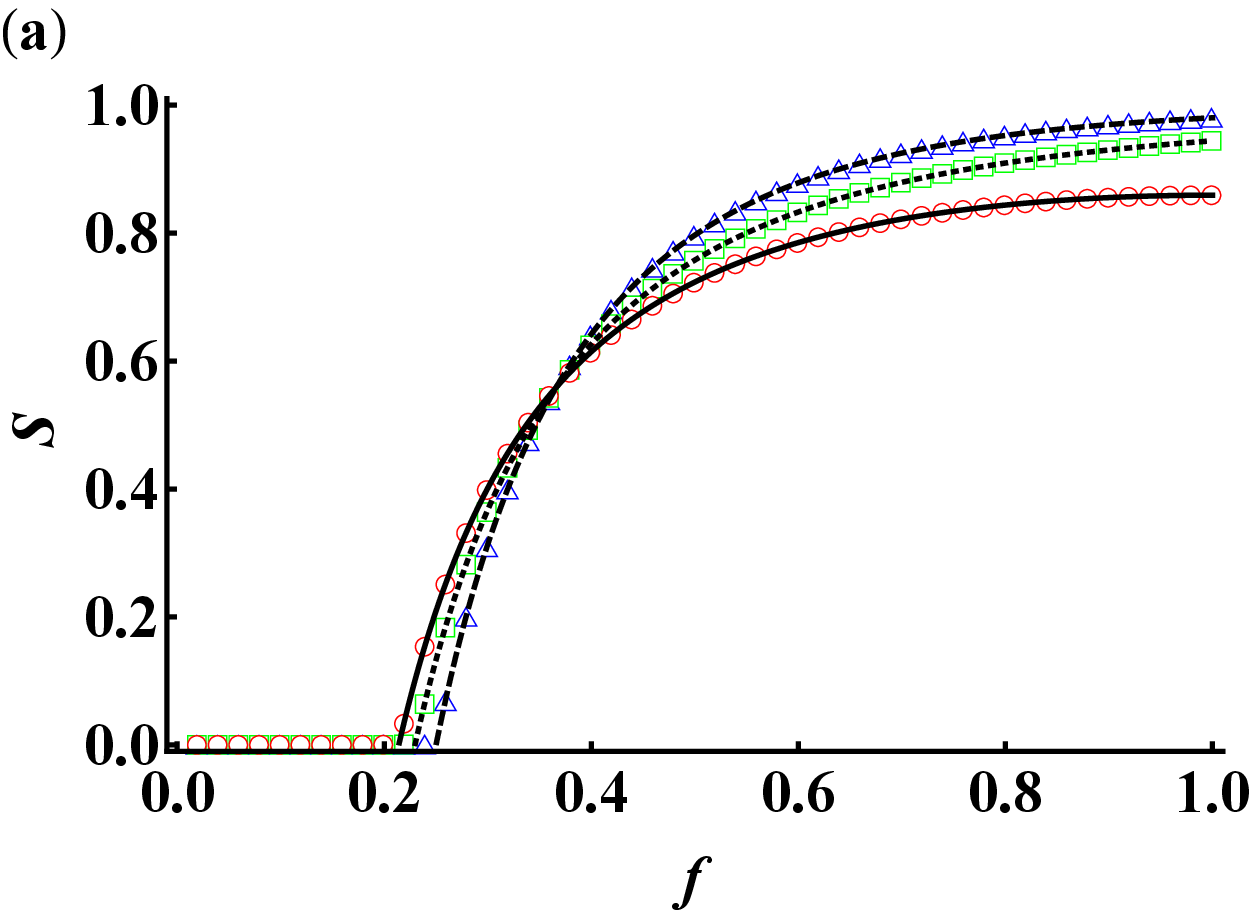}
\includegraphics[height=.23\textwidth]{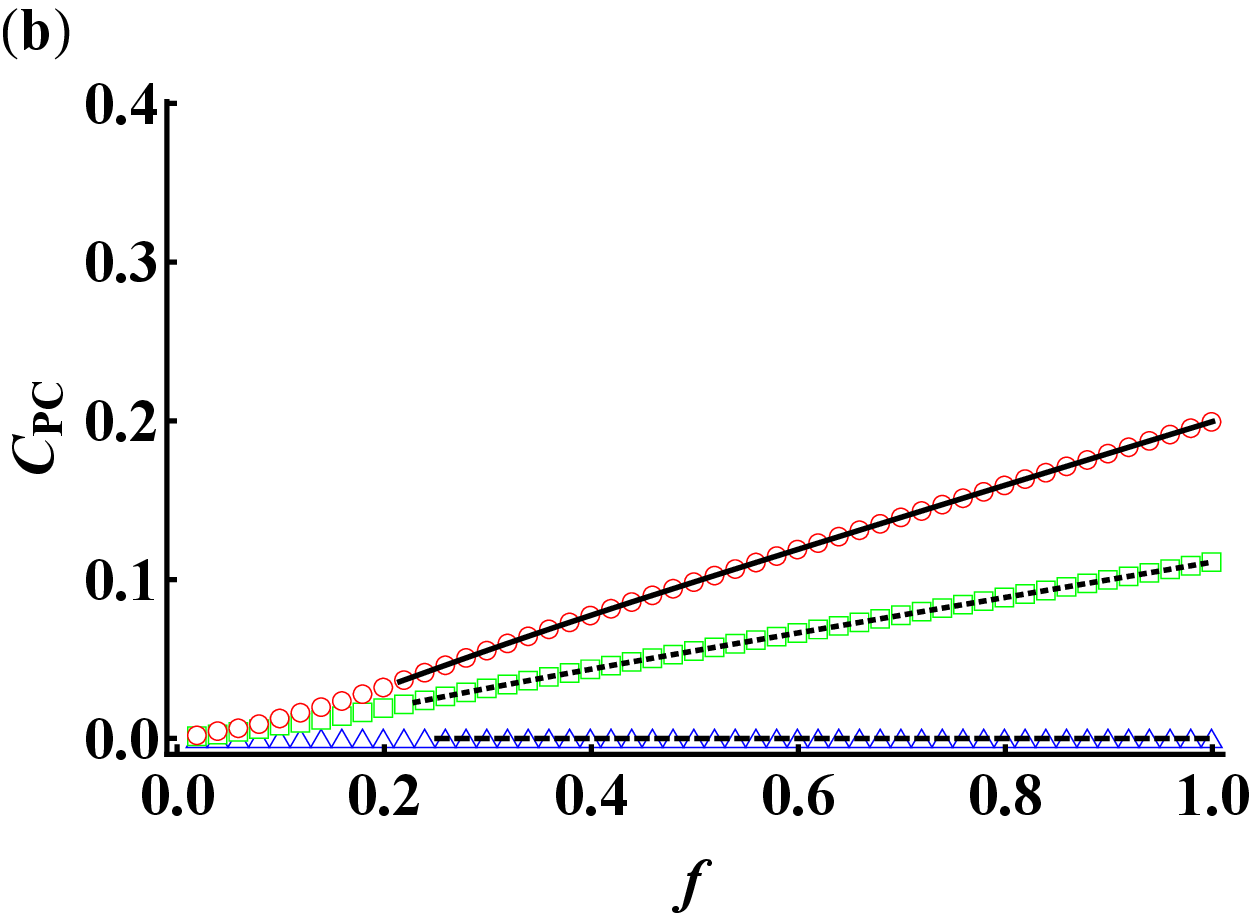}
\includegraphics[height=.23\textwidth]{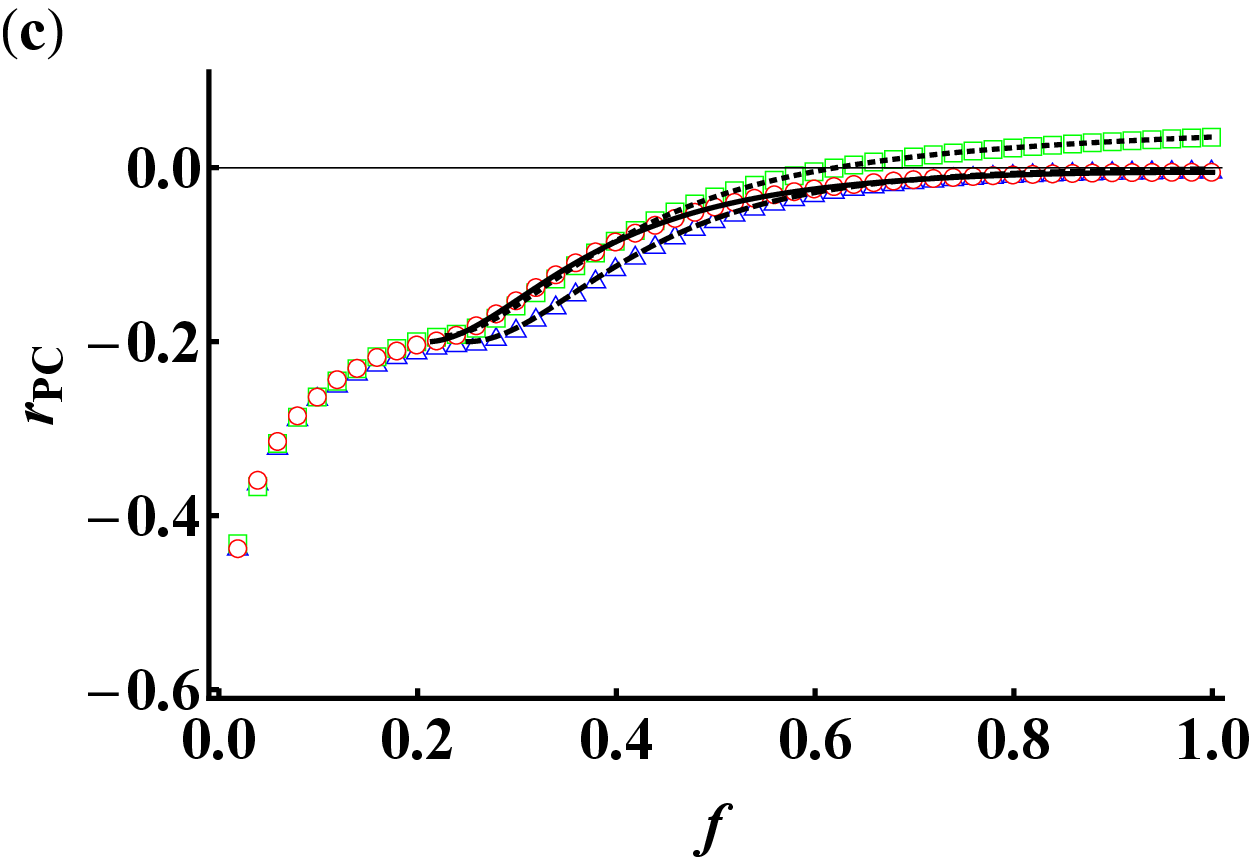}
\caption{
Comparison with the analytical treatments and simulation results for the structures of the PC formed by bond percolation: (a) normalized size $S$, (b) clustering coefficient $C_{\rm PC}$, and (c) assortative coefficient $r_{\rm PC}$. 
The Poisson RCN with $\langle k \rangle=4$ is utilized as the original network. 
The simulation results are for the cases of $\langle s \rangle=0$ and $\langle t \rangle=2$ (red circles), $\langle s \rangle=2$ and $\langle t \rangle=1$ (green squares), and $\langle s \rangle=4$ and $\langle t \rangle=0$ (blue triangles).
The corresponding analytical estimates, whose derivations are given in Appendix~\ref{sec:appendix}, are represented by the solid lines, dotted lines, and dashed lines.
}
\label{fig:poissonRCN:bond}
\end{center}
\end{figure}

Figures~\ref{fig:poissonRCN:site} (a)--(c) show the $f$ dependence of the normalized PC size, $S$, the clustering coefficient of the PC, $C_{\rm PC}$, and the assortative coefficient of the PC, $r_{\rm PC}$, respectively, for site percolation on the Poisson RCN with $\langle k \rangle=4$ and several combinations of $\langle s \rangle$ and $\langle t \rangle$.
The solid, dotted, and dashed lines given only for $f>f_c$ represent the analytical estimates and the symbols (red circles, green squares, and blue triangles) represent the Monte Carlo results.
Our analytical estimates perfectly matched with the simulation results for $f > f_{\rm c}$ in all cases.
As shown in Fig.~\ref{fig:poissonRCN:site} (a), the PC on the Poisson RCN emerges at $f_c=1/4$, irrespective of the value of $C_0$ \footnote{For Poisson RCNs where $G_p(x,y)=G_q(x,y)=G_r(x,y)$, from $\det|A-I|=0$, $f_c(\langle s \rangle + 2 \langle t \rangle) = f_c \langle k \rangle =1$ is obtained. 
Thus, the percolation threshold of site percolation is $f_c=1/\langle k \rangle$, which depends only on the average degree $\langle k \rangle$.}. 
(Note that for bond percolation $f_c$ depends on the value of $C_0$ \footnote{Bond percolation on the RCNs has already been examined \cite{newman2009random,miller2009percolation,gleeson2010clustering}. 
Newman \cite{newman2009random} demonstrated that for the Poisson RCN, the clustering decreases the bond percolation threshold $f_c$. 
On the other hand, by comparing $f_c$ for clustered networks and unclustered networks of the same degree correlation, Miller \cite{miller2009percolation} clarified that this decrease in $f_c$ is not caused by the influence of the clustering, but by the influence of the assortative correlation specific to the Poisson RCN. 
Furthermore, Gleeson et al. \cite{gleeson2010clustering} arrived at the same conclusion by using their clustered $\gamma$-theory networks.
Thus, the network clustering increases $f_c$ for the delta RCN, which has no degree-degree correlation, although it decreases $f_c$ for the Poisson RCN.
}, as shown in Fig.~\ref{fig:poissonRCN:bond} (a).)
Figures~\ref{fig:poissonRCN:site} (b) and (c) show that both the clustering coefficient $C_{\rm PC}$ and the assortative coefficient $r_{\rm PC}$ of the PC exhibit no singular behaviors at and around the percolation threshold $f_c$. 
Moreover, we notice that $C_{\rm PC}>0$ at $f=f_c$ (i.e., the PC is already highly clustered when it emerged), although it is a fractal (i.e., a fractal PC).
For the case of $\langle t \rangle =0$, the assortative coefficient, $r_{\rm PC}$, of the PC is always negative (see the blue triangles in Fig.~\ref{fig:poissonRCN:site} (c)) and takes $-1/5$ at $f=f_c$, as already derived in~\cite{mizutaka2018disassortativity}.
The assortative coefficient of the PC becomes positive for a large $f$ if the original RCN is assortative (see the red circles and green squares in Fig.~\ref{fig:poissonRCN:site} (c)), although $r_{\rm PC}$ always becomes negative at and around $f_c$, irrespective of the assortativity of the original network. 
In Figs.~\ref{fig:poissonRCN:site:diagram} (a) and (b), we plot the clustering coefficient and assortative coefficient of the fractal PC, i.e., $C_{\rm PC}$ and $r_{\rm PC}$ at $f=f_c$, in the ($\langle s \rangle$, $\langle t \rangle$) plane.
The clustering coefficient of the fractal PC becomes smaller at larger $\langle s \rangle$ and larger $\langle t \rangle$ (Fig.~\ref{fig:poissonRCN:site:diagram} (a)).
Figures~\ref{fig:poissonRCN:site:diagram} (b) and (c) clearly show that the assortative coefficient of the fractal PC is negative for any $\langle s \rangle$ and $\langle t \rangle$.
This supports the disassortativity of the fractal PC, according with our previous work~\cite{mizutaka2018disassortativity} for uncorrelated networks.

The analysis of the RCN, however, does not necessarily give the disassortativity of fractal PCs. 
Figure~\ref{fig:deltaRCN:site} shows the results of $S$, $C_{\rm PC}$, and $r_{\rm PC}$, for the delta RCN of $s_0+t_0=4$.
We again observe that $C_{\rm PC}$ and $r_{\rm PC}$ show no singular behaviors at $f_c$ and $C_{\rm PC}>0$ already at $f=f_c$ (if $C_0>0$).
However, both analytical estimate and simulations yield a different conclusion as regards the assortativity: $r_{\rm PC}>0$ even at $f_c$ if the delta RCN is clustered (see the red circles and green squares in Fig.~\ref{fig:deltaRCN:site} (c)). 
In Fig.~\ref{fig:deltaRCN:site:diagram}, we calculate $C_{\rm PC}$ and $r_{\rm PC}$ at $f=f_c$ for the delta RCNs with changing $s_0$ and $t_0$ in the range $k=3$--$12$, and display their scatter plot.
Figures~\ref{fig:deltaRCN:site:diagram} (a) and (b) show that $C_{\rm PC}$ and $r_{\rm PC}$ of the fractal PC tend to decrease as degree $k$ increases. 
Furthermore, in the delta RCNs with a fixed value of $k$, $C_{\rm PC}$ and $r_{\rm PC}$ of the fractal PC decrease as the number of triangles per node $t_0$ decreases.
As the blue circles in Fig.~\ref{fig:deltaRCN:site:diagram} (c) indicate, the fractal PC in the delta RCN can be assortative when it is highly clustered or, equivalently, the original delta RCN has small $k$ and large $C_0$.

\begin{figure}[!b]
\begin{center}
(a)
\includegraphics[width=.36\textwidth]{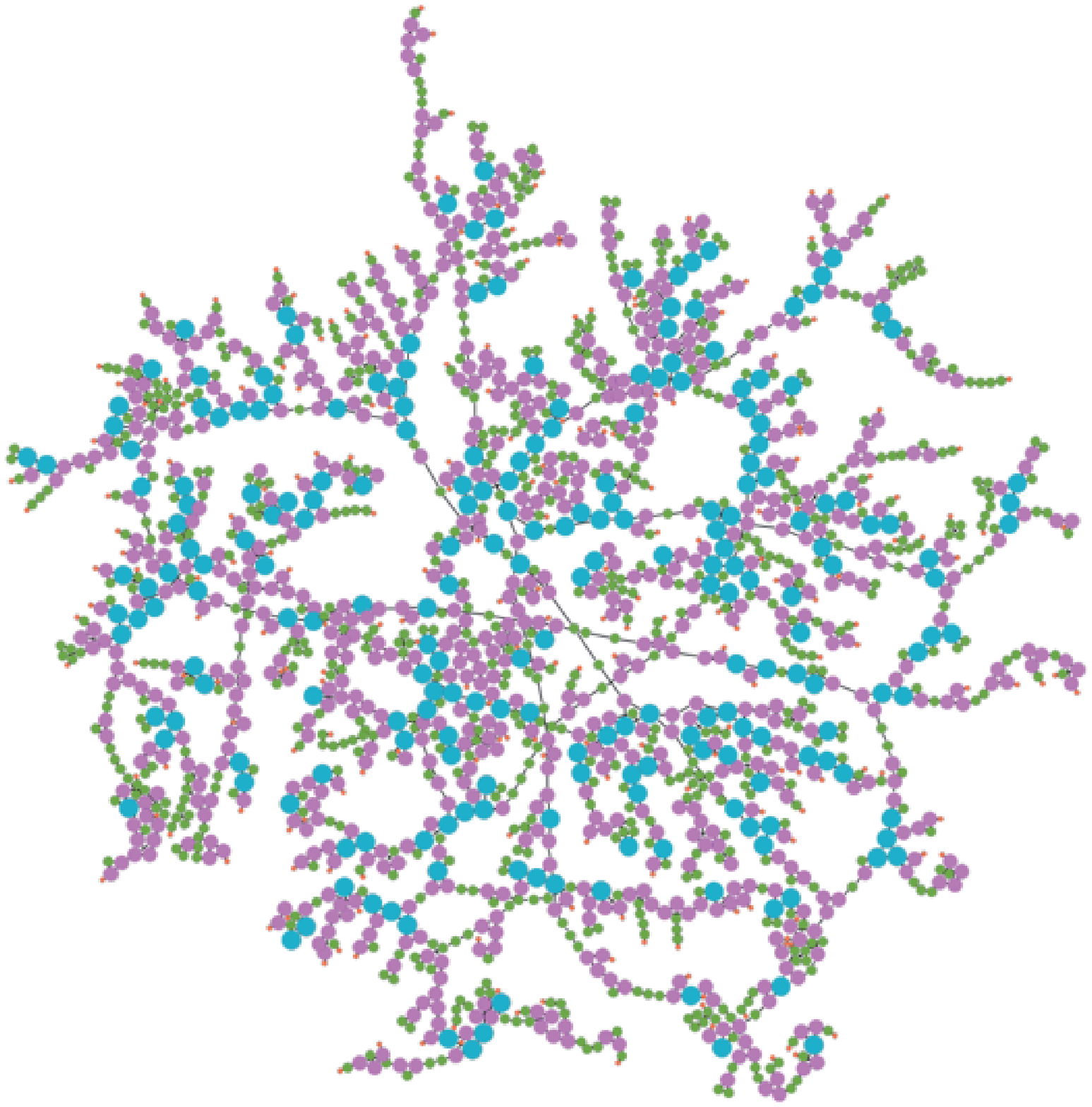}
\hspace{0.5cm}
(b)
\includegraphics[width=.32\textwidth]{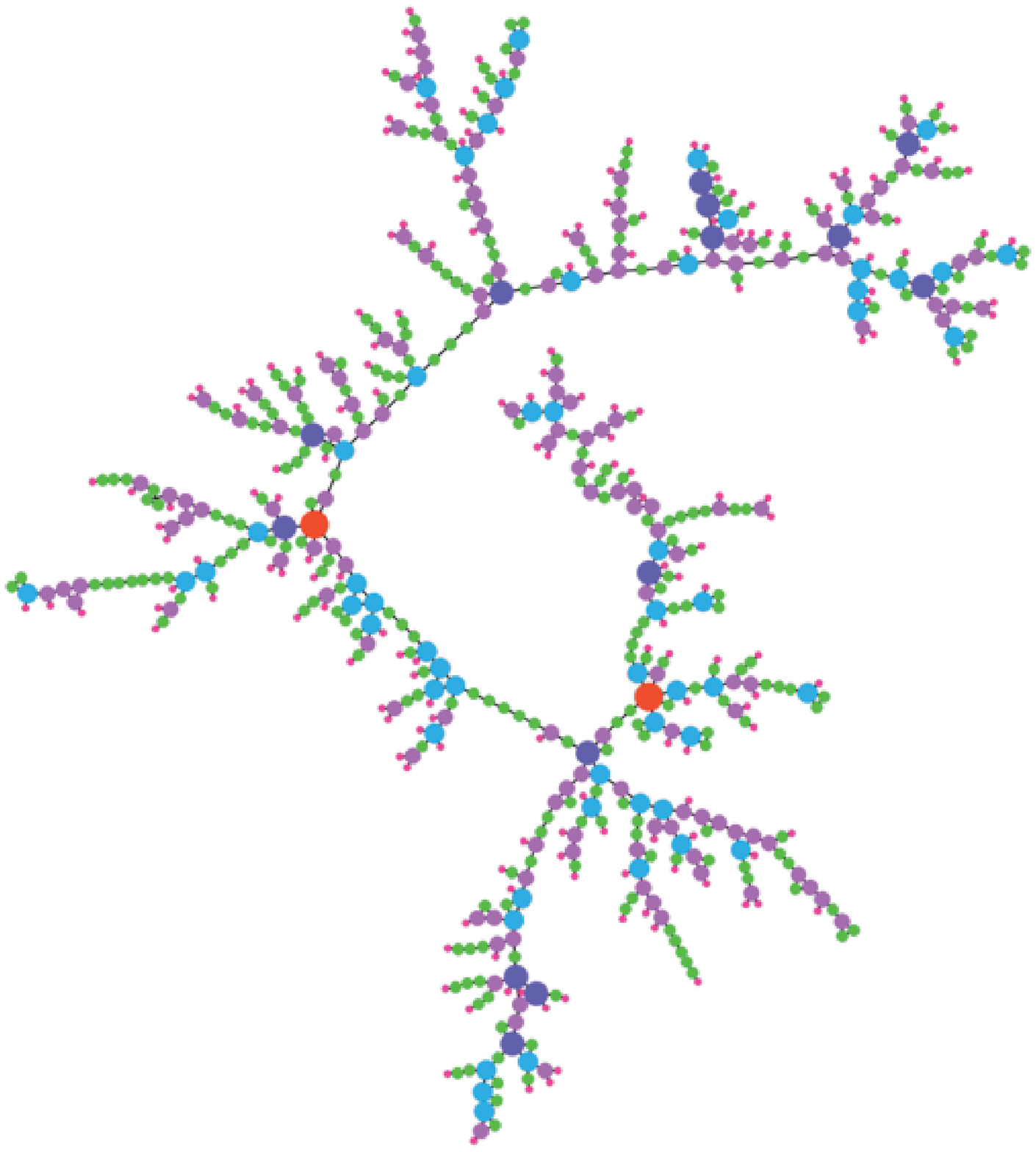}
\caption{
Snapshot of (a) the largest cluster at the percolation threshold and (b) the renormalized one using the box counting scheme with $l_B=4$. 
The numbers of nodes and triangles decrease from 1913 and 480 in (a) to 509 and 37 in (b), respectively.
The assortative coefficient of the largest cluster is then changed from 0.215 to $-$0.126.
}
\label{fig:criticalInfiniteClsuters}
\end{center}
\end{figure}

\begin{figure}[!h]
\begin{center}
\includegraphics[height=.23\textwidth]{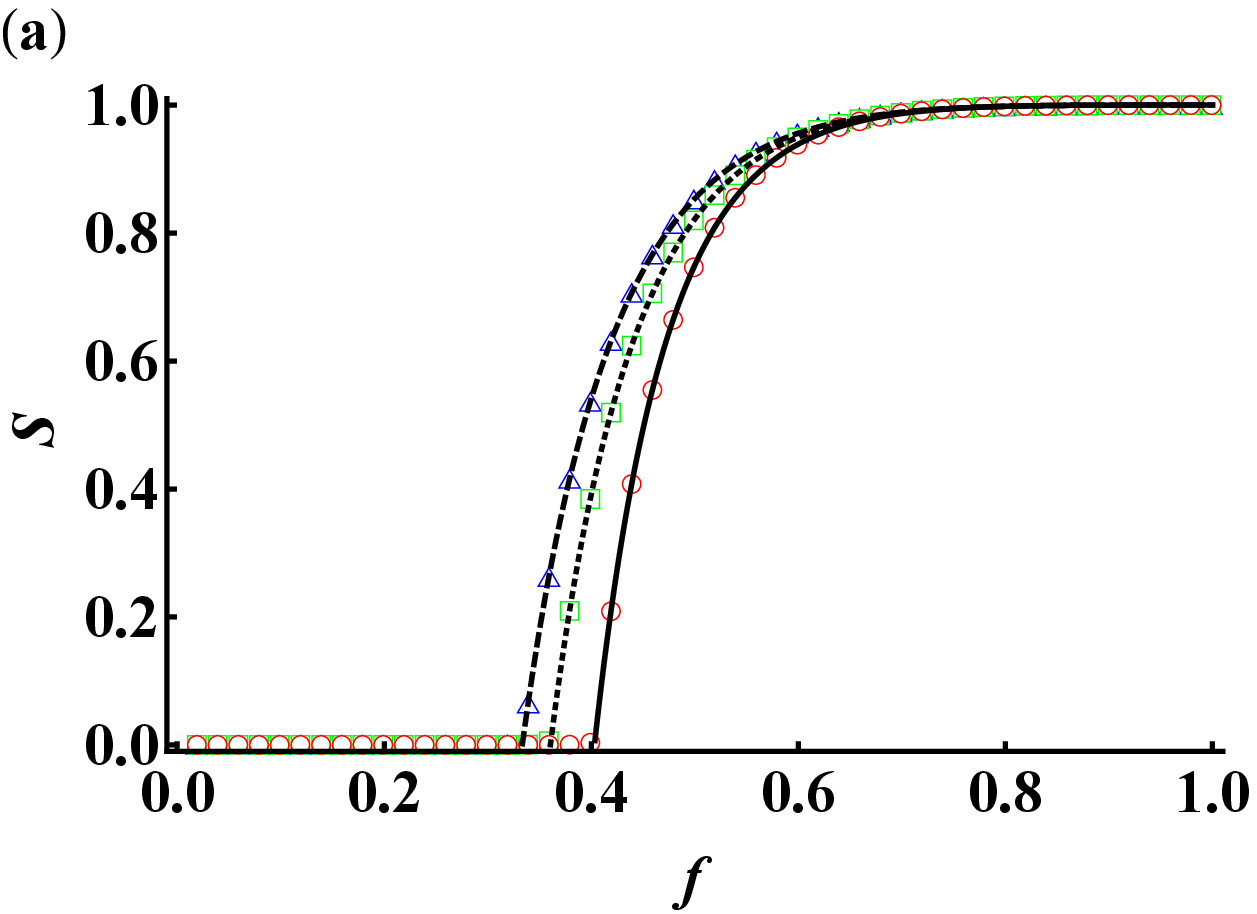}
\includegraphics[height=.23\textwidth]{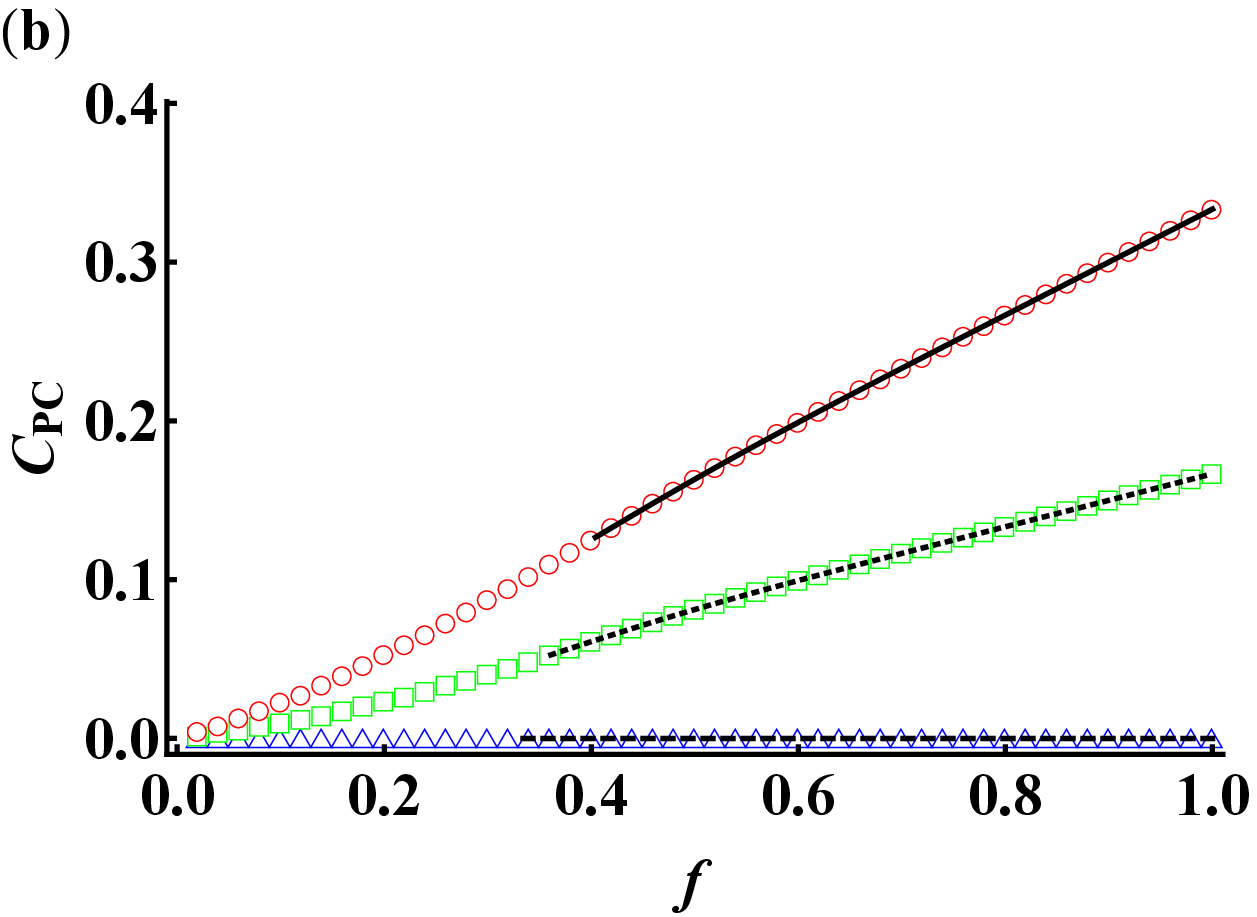}
\includegraphics[height=.23\textwidth]{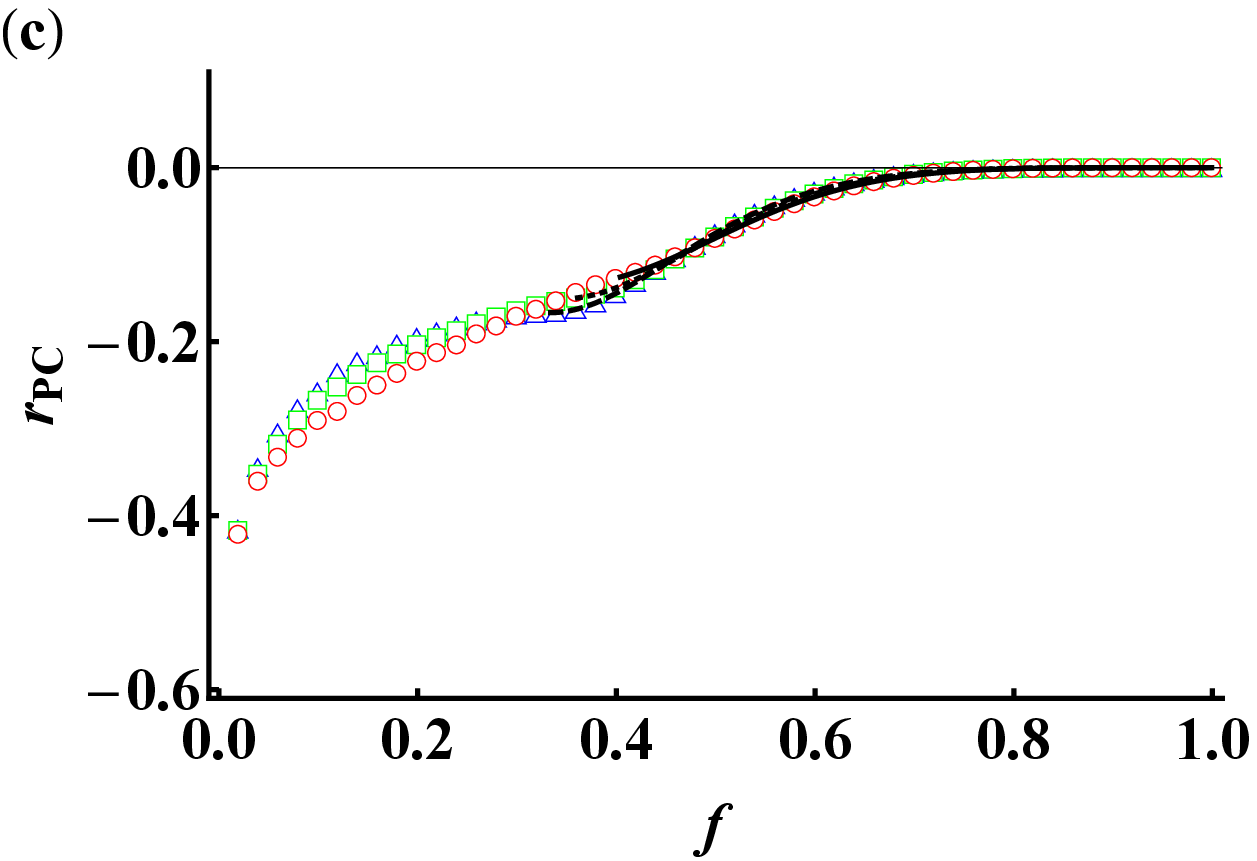}
\caption{
Comparison with the analytical treatments and simulation results for the structures of the PC formed by bond percolation: (a) normalized size $S$, (b) clustering coefficient $C_{\rm PC}$, and (c) assortative coefficient $r_{\rm PC}$. 
The delta RCN with degree 4 is utilized as the original network. 
The simulation results are for the cases of $s_0=0$ and $t_0=2$ (red circles), $s_0=2$ and $t_0=1$ (green squares), and $s_0=4$ and $t_0=0$ (blue triangles).
The corresponding analytical estimates, whose derivations are given in Appendix~\ref{sec:appendix}, are represented by the solid lines, dotted lines, and dashed lines.
}
\label{fig:deltaRCN:bond}
\end{center}
\end{figure}

\section{Discussion}\label{sec:discussion}

In this study, we derived the clustering coefficient $C_{\rm PC}$ and assortative coefficient $r_{\rm PC}$ of the percolating cluster (PC) formed by site percolation on the random clustered network (RCN), thereby validating the disassortativity of fractal networks.
Applying our formulation to the RCN whose joint probability of single edges and triangles obeys a double Poisson distribution (Poisson RCN) and the RCN whose nodes have the same numbers of single edges and triangles (delta RCN), we confirmed that our analytical estimates for $C_{\rm PC}$ and $r_{\rm PC}$ perfectly agree with the simulation results.
Our results signified that both the clustering coefficient and the assortative coefficient of the PC do not exhibit any singular behavior near the percolation threshold, and the PC at the percolation threshold, namely, the fractal PC, is clustered as long as an underlying RCN is clustered.
As regards the assortativity of the PC, the result seemingly contradicts the disassortativity of the fractal networks: the fractal PCs exhibit $r_{\rm PC}<0$ for all Poisson RCNs and most delta RCNs, but $r_{\rm PC}>0$ for only a few delta RCNs.

The question remains as regards to whether the last result immediately denies the disassortativity of the fractal networks.
We should note that the positive assortativity of the delta RCN is easily lost.
For example, we revisit site percolation on the delta RCN with $p_{s,t}=\delta_{s,0}\delta_{t,2}$, in which $r_{\rm PC}>0$ for a fractal PC [red circles in Fig.~\ref{fig:deltaRCN:site} (c)].
This network consists of only triangles; triangles are a basic unit giving a characteristic scale.
Let us consider applying the box covering scheme \cite{song2005self, song2006origins} to a fractal PC formed on this network [Fig.~\ref{fig:criticalInfiniteClsuters} (a)].
Tiling a fractal PC with the estimated minimum number of boxes of a linear size $l_B=4$ and renormalizing it so that each box is replaced as a supernode [Fig.~\ref{fig:criticalInfiniteClsuters} (b)], we recalculated the clustering coefficient and the assortative coefficient of the renormalized ones.
Renormalization breaks the characteristic scale (triangle) and unveils a disassortative structure: the clustering coefficient and the assortative coefficient are changed from $C_{\rm PC} \approx 0.333$ and $r_{\rm PC} \approx 0.245$ to $C_{\rm PC} \approx 0.112$ and $r_{\rm PC} \approx -0.129$, respectively, under renormalization \footnote{
To take the average, we simulated site percolation of $f=f_c$ 10 times on each of the 10 network realizations, picked up the largest clusters, and applied the box cover scheme of $l_{B}=4$ 10 times to each largest cluster.
The number of nodes in the delta RCN was $90000$. 
The average number of nodes in the largest clusters was changed from 2571 to 675 under renormalizations.}.
It indicates that the assortativity of a fractal PC in the delta RCN is attributed to the characteristic scale of the triangles. 
The fractal PC formed by site percolation on the delta RCN appears disassortative for larger scales.
As was shown in Fig.~\ref{fig:deltaRCN:site:diagram}, the assortativity of the fractal PC is observed in only 5 combinations of $s_0$ and $t_0$, i.e., $(s_0, t_0) = (1,1), (2,1), (0,2), (1,2), (0,3)$. 
In these combinations, every node has equally few edges, most of which form triangles.
As with the present example ($p_{s,t}=\delta_{s,0}\delta_{t,2}$), it is likely that the assortativity observed for delta RCNs is attributed to the triangles giving a characteristic scale and is easily broken by rescaling.

Moreover, the disassortativity of a fractal PC arises in the delta RCN when it is formed by {\it bond percolation}. 
Similar to site percolation, we can derive the clustering coefficient and the assortative coefficient of the PC formed by bond percolation on the RCN (see Appendix~\ref{sec:appendix}).
Figures~\ref{fig:poissonRCN:bond} and \ref{fig:deltaRCN:bond} show the structural properties of the PC formed by bond percolation on the Poisson and delta RCNs, respectively.
The triangles are easily broken in bond percolation in contrast to the formation of the PC: the clustering coefficient of the fractal PC is relatively small when compared with site percolation [Figs.~\ref{fig:poissonRCN:bond} (b) and ~\ref{fig:deltaRCN:bond} (b)].
As indicated by both analytical estimates and simulation results, $r_{\rm PC}$ is negative at $f=f_c$ for not only the Poisson RCN but also for the delta RCN [Figs.~\ref{fig:poissonRCN:bond} (c) and ~\ref{fig:deltaRCN:bond} (c)].
The fractal PC is thus disassortative on the delta RCN when formed by bond percolation.

Having considered these results, it can be presumed that the fractal networks formed by percolation processes on networks are disassortative in essence. 
Further studies on the disassortativity of fractal networks should be conducted.
This study has concentrated on the degree-degree correlation of networks, but has not asked for higher-order degree correlations, such as the long-range degree correlation \cite{rybski2010quantifying,fujiki2018general,mayo2015long,orsini2015quantifying}.
Our formulation in this study may be extended to compute the long-range correlation, and it is expected to be provided by future work.

\section*{Acknowledgements}
S.M.\ and T.H.\ acknowledge financial support from JSPS (Japan) KAKENHI Grant Number JP18KT0059.
S.M.\ was supported by a Grant-in-Aid for Early-Career Scientists (No.~18K13473) and a Grant-in-Aid for JSPS Research Fellow (No.~18J00527) from the Japan Society for the Promotion of Science (JSPS) for performing this work.
T.H.\ acknowledges financial support from JSPS (Japan) KAKENHI Grant Numbers JP16H03939 and JP19K03648.

\appendix

\section{Structure of the percolating cluster formed by bond percolation in random clustered network}\label{sec:appendix}

\begin{figure}[b]
\begin{center}
\includegraphics[width=.70\textwidth]{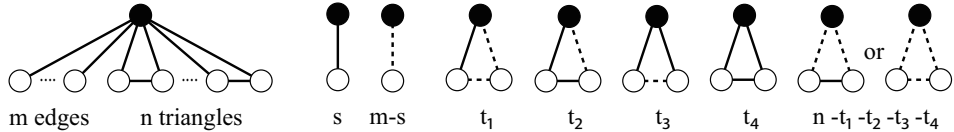}
\caption{
Motifs of the single edges and triangles in a cluster formed by bond percolation.
The solid and dashed lines represent the open and closed edges, respectively.
}
\label{fig:schematic}
\end{center}
\end{figure}

In this Appendix, we treat bond percolation on the RCN: each edge is open (not removed) with probability $f$ and closed (removed from the original network) otherwise (probability $\tf=1-f$).
The normalized PC size, $S$, is given by solving the following equations \cite{newman2009random}: 
\be
S=1-G_p(u,v^2),
\ee
with
\be
u=\tf + f G_q(u,v^2) \label{eq:recUbond}
\ee
and
\begin{eqnarray}
v^2 &=& 1-2f \tf^2 (1-G_r(u,v^2))-f^2 (3\tf+f)(1-G_r(u,v^2)^2) \nonumber \\
&=& \tf^2+2\tf^2 f G_r(u,v^2) +f^2(3\tf+f)G_r(u,v^2)^2. \label{eq:recVbond}
\end{eqnarray}
Here $v^2$ is the probability that the two adjacent nodes forming a triangle with a node are not members of the PC.
The percolation threshold $f_c$ is given as the point above which $u<1$ and $v<1$ are the solution of the above-mentioned equations.
By linearizing Eqs.~(\ref{eq:recUbond}) and (\ref{eq:recVbond}) around $(u,v)=(1,1)$ and examining the condition $\det|A-I|=0$, where 
\be
A=
\left[
\begin{array}{cc}
f\sum s q_{s,t} & f\sum t q_{s,t} \\
2f(1+f\tf) \sum s r_{s,t} & 2f(1+f\tf) \sum t r_{s,t}
\end{array}
\right]
\label{eq:matrixBond}
\ee
and $I$ is the identity matrix, we obtain the percolation threshold $f_c$.
We hereafter assume $f>f_c$, focusing on the PC. 

First, we derive the clustering coefficient of the PC. 
A randomly chosen node has $m$ single edges and $n$ triangles in an original network with probability $p_{m,n}$.
We consider the probability $P({\rm PC}, s, t_1, t_2, t_3, t_4)$ that a randomly chosen node belongs to the PC and has $s$ single edges and $t_1, t_2, t_3$, and $t_4$ motifs (shown in Fig.~\ref{fig:schematic}) in the PC. 
This probability is presented as follows:
\begin{eqnarray}
P({\rm PC}, s, t_1, t_2, t_3, t_4) &=& 
\sum_{m,n} p_{m,n} \binom{m}{s} f^s \tf^{m-s} \binom{n}{t_{1},t_{2},t_{3},t_{4}} \\
&&\quad\quad\quad \times
 f^{3t_4} (f^2 \tf)^{t_3} (2f^2 \tf)^{t_2} (2f \tf^2)^{t_1} \tf^{2(n-t_1-t_2-t_3-t_4)} (1-\tu^{s}\tv^{t_1+2t_2+2t_3+2t_4}),
\nonumber
\end{eqnarray}
where $\tu=G_q(u,v^2)$ and $\tv=G_r(u,v^2)$.
We denote by $F_{\rm PC}(x,y_1,y_2,z_1,z_2)$ the generating function for the probability $P_{\rm PC}(s, t_1, t_2, t_3, t_4)$ that a node randomly chosen from the PC has $s$ single edges and $t_1, t_2, t_3$, and $t_4$ motifs in Fig.~\ref{fig:schematic}. 
Using $P_{\rm PC}(s, t_1, t_2, t_3, t_4)=P({\rm PC}, s, t_1, t_2, t_3, t_4)/P({\rm PC})$, where $P({\rm PC})$ is the probability that a randomly chosen node is a member of the PC, 
\begin{eqnarray}
P({\rm PC})=1-G_p(u,v^2)=S,
\end{eqnarray}
and after some transformations, we obtain
\begin{eqnarray}
F_{\rm PC}(x,y_1,y_2,z_1,z_2)
&=& \sum_{s=0}^\infty \sum_{t_1=0}^\infty \sum_{t_2=0}^\infty \sum_{t_3=0}^\infty \sum_{t_4=0}^\infty P_{\rm PC}(s, t_1, t_2, t_3, t_4) x^s y_1^{t_1} y_2^{t_2} z_1^{t_3} z_2^{t_4}
\nonumber \\
&=&\frac{1}{P({\rm PC})}G_p(\tf+f x, \tf^2 +2\tf^2 f y_1 +2\tf f^2 y_2 + \tf f^2 z_1+f^3 z_2) \\
&&-\frac{1}{P({\rm PC})}G_p(\tf+f \tu x, \tf^2 +2 \tf^2 f \tv y_1 + 2\tf f^2\tv^2 y_2+\tf  f^2 \tv^2 z_1+f^3 \tv^2 z_2). \nonumber
\end{eqnarray}
The degree distribution of the PC, $P_{\rm PC}(k)$, is given as
\be
P_{\rm PC}(k) = \frac{1}{k!} \frac{\partial^k}{\partial x^k} F_{\rm PC}(x,x,x,x^2,x^2) \Big|_{x=0},
\ee
and the clustering coefficient of the PC, $C_{\rm PC}$, is given as $C_{\rm PC}=3 N_\Delta/N_3$, where 
\be
3 N_\Delta = \frac{\partial}{\partial z_2} F_{\rm PC}(x,y_1,y_2,z_1,z_2)\Big|_{x=y_1=y_2=z_1=z_2=1},
\ee
and
\be
N_3 = \frac{1}{2} \frac{\partial^2}{\partial x^2} F_{\rm PC}(x,x,x,x^2,x^2)\Big|_{x=1}.
\ee

Next, we formalize the assortative coefficient of the PC, $r_{\rm PC}$. 
The derivation of $r_{\rm PC}$ for bond percolation is the same as that for site percolation in Sec.~\ref{sec:assortativity}.

The generating function for $Q({\rm PC}, s_1, t_1, s_2, t_2)$, which is the probability that an edge belongs to the PC and its ends have $s_1$ single edges and $t_1$ triangle edges and $s_2$ single edges and $t_2$ triangle edges except the selected edge in the PC, respectively, is
\be
H(x_1,x_2,y_1,y_2)=P_s H_s(x_1,x_2,y_1,y_2)+P_t H_t (x_1,x_2,y_1,y_2),
\ee
where
\begin{eqnarray}
H_s(x_1,x_2,y_1,y_2)= 
f G_q(g_1(x_1),g_2(x_2)) G_q(g_1(y_1),g_2(y_2)) - f G_q(h_1(x_1),h_2(x_2))G_q(h_1(y_1),h_2(y_2)) \label{eq:Hs:bond}
\end{eqnarray}
and
\begin{eqnarray}
H_t(x_1,x_2,y_1,y_2)
&=& 
f (\tf^2+\tf f (x_2+y_2)+f^2 x_2y_2) G_r(g_1(x_1),g_2(x_2)) G_r(g_1(y_1),g_2(y_2))
\nonumber \\ && 
- f (\tf^2+\tf f \tv (x_2+y_2)+f^2 \tv x_2y_2) G_r(h_1(x_1),h_2(x_2))G_r(h_1(y_1),h_2(y_2)). \label{eq:Ht:bond}
\end{eqnarray}
We used the following notations in Eqs.~(\ref{eq:Hs:bond}) and (\ref{eq:Ht:bond}):
\be
g_1(x)=\tf+f x, \; 
g_2(x)=\tf^2 +2\tf f x+f^2 x^2
\ee
and
\be
h_1(x)=\tf+f \tu x, \;
h_2(x)=\tf^2 +2\tf^2 f \tv x+2\tf f^2 \tv^2 x+f^2 \tv^2 x^2.
\ee
The generating function for the probability $Q({\rm PC}, k, k')$ that an edge belongs to the PC and has two ends with degrees $k+1$ and $k'+1$ in the PC is then presented as $\sum_{k,k'} Q({\rm PC}, k, k') x^{k} y^{k'} =H(x,x,y,y)$ 
and the generating function for the probability $Q({\rm PC}, k)=\sum_{k'}Q({\rm PC}, k, k')$ that an edge belongs to the PC and reaches a node with degree $k+1$ is given as $\sum_{k} Q({\rm PC}, k) x^{k} =H(x,x,1,1)$.
The probability that a randomly chosen edge is open and belongs to the PC is 
\begin{eqnarray}
Q({\rm PC})&=&H(1,1,1,1) \nonumber \\
&=&f P_s (1-\tu^2)+f P_t (1-(\tf^2+f(2\tf+f)\tv)\tv^2).
\end{eqnarray}
Then the assortative coefficient of PC, $r_{\rm PC}$, is given as
\be
r_{\rm PC}=\frac{\partial_x \partial_y B_{\rm PC}(x,y) - (\partial_x S_{\rm PC}(x))^2}{(x \partial_x)^2 S_{\rm PC}(x) - (\partial_x S_{\rm PC}(x))^2} \Big|_{x=y=1},
\ee
where $B_{\rm PC}(x,y)$ is the generating function for $Q_{\rm PC}(k, k')=Q({\rm PC}, k, k')/Q({\rm PC})$,
\begin{eqnarray}
B_{\rm PC}(x,y)
&=& \frac{H(x,x,y,y)}{Q({\rm PC})} \nonumber \\
&=& \frac{f P_s}{Q({\rm PC})} \Big[G_q(g_1(x),g_2(x)) G_q(g_1(y),g_2(y))-G_q(h_1(x),h_2(x)) G_q(h_1(y),h_2(y)) \Big]
\nonumber \\
&&+\frac{f P_t}{Q({\rm PC})} \Big[ (\tf^2+\tf f (x+y)+f^2 xy) G_r(g_1(x),g_2(x)) G_r(g_1(y),g_2(y))
\nonumber \\ && \quad
- (\tf^2+\tf f \tv (x+y)+f^2 \tv xy) G_r(h_1(x),h_2(x))G_r(h_1(y),h_2(y)) \Big],
\end{eqnarray}
and $S_{\rm PC}(x)$ is the generating function for $Q_{\rm PC}(k)=\sum_{k'}Q_{\rm PC}(k, k')$, 
\begin{eqnarray}
S_{\rm PC}(x) 
&=& \frac{H(x,x,1,1)}{Q({\rm PC})} \nonumber \\
&=& \frac{f P_s}{Q({\rm PC})} \Big[G_q(g_1(x),g_2(x))-G_q(h_1(x),h_2(x)) \tu \Big]
\\&&
+\frac{f P_t}{Q({\rm PC})} \Big[ (\tf+f x) G_r(g_1(x),g_2(x))-(\tf^2+\tf f \tv +f \tv x) G_r(h_1(x), h_2(x)) \tv \Big].
\nonumber
\end{eqnarray}


\end{document}